\documentclass[a4paper,11pt]{article}
\pdfoutput=1 

\usepackage{jheppub} 

\usepackage[T1]{fontenc} 

\def\tr{\text{tr}}

\def\hat{\widehat}

\def\del{{\partial}}

\def\a{{\alpha}}
\def\b{{\beta}}

\def\G{{\Gamma}}

\def\D{{\Delta}}

\def\h{{\eta}}
\def\th{{\theta}}

\def\l{{\lambda}}
\def\L{{\Lambda}}
\def\m{{\mu}}
\def\n{{\nu}}

\def\s{{\sigma}}
\def\t{{\tau}}
\def\f{{\varphi}}
\def\o{{\omega}}

\def\Z{{\mathbb Z}}

\usepackage{verbatim}   
\usepackage[all]{xy}
\usepackage{bm}  
\def\Dslash{{\rlap{\raise 1pt \hbox{$\>/$}}D}}
\def\Pslash{{\rlap{\raise  1pt \hbox{$\>/$}}\,\partial}}

\newcommand{\diff}{\mathrm{d}}

\newcommand{\Diff}{{\mathcal{D}}}

\newcommand{\im}{\mathrm{i}}

\newcommand{\rme}{\mathrm{e}}

\def\l{\ell}

\def\tr{{\rm tr}}

\def\Z{{\mathbb Z}}

\usepackage{comment}

\newcommand{\dd}{\text{d}}
\newcommand{\ii}{\text{i}}
\newcommand{\ee}{\text{e}}
\newcommand{\st}{\Tilde{s}}
\newcommand{\lt}{\Tilde{\ell}}

\newcommand{\xt}{\Tilde{x}}

\usepackage{bm}
\usepackage{sansmath}

\preprint{YITP-21-01}

\title{\boldmath Semi-Abelian gauge theories, non-invertible symmetries, and string tensions beyond $N$-ality}

\author[1]{Mendel Nguyen,}
\affiliation[1]{Department of Physics, North Carolina State University, Raleigh, NC 27607, USA}
\author[2]{Yuya Tanizaki,} 
\affiliation[2]{Yukawa Institute for Theoretical Physics, Kyoto University, Kyoto 606-8502, Japan}
\author[1]{Mithat \"Unsal}
\emailAdd{mendelnguyen@gmail.com, yuya.tanizaki@yukawa.kyoto-u.ac.jp, unsal.mithat@gmail.com}

\abstract{
We study a $3$d lattice gauge theory with gauge group $\mathrm{U}(1)^{N-1}\rtimes \mathrm{S}_N$, which is obtained by gauging the $\mathrm{S}_N$ global symmetry of a pure $\mathrm{U}(1)^{N-1}$ gauge theory, and we call it the semi-Abelian gauge theory. 
We compute mass gaps and string tensions for both theories using the monopole-gas description. 
We find that the effective potential receives equal contributions at leading order from monopoles associated with the entire $\mathrm{SU}(N)$ root system. 
Even though the center symmetry of the semi-Abelian gauge theory is given by $\mathbb{Z}_N$, we observe that the string tensions do not obey the $N$-ality rule and carry more detailed information on the representations of the gauge group. 
We find that this refinement is due to the presence of non-invertible topological lines as a remnant of $\mathrm{U}(1)^{N-1}$ one-form symmetry in the original Abelian lattice theory. 
Upon adding charged particles corresponding to $W$-bosons, such non-invertible symmetries are explicitly broken so that the $N$-ality rule should emerge in the deep infrared regime.
}

\begin{document} 
\maketitle
\flushbottom

\section{Introduction}
 
    In all calculable, confining $\mathrm{SU}(N)$ gauge theories in continuum, such as the Polyakov model on $\mathbb{R}^3$ \cite{Polyakov:1975rs, Polyakov:1987ez}, the Seiberg--Witten model on $\mathbb{R}^4$ \cite{Seiberg:1994rs}, and deformed Yang--Mills and adjoint QCD on $\mathbb{R}^3 \times S^1$ \cite{Unsal:2008ch, Unsal:2007jx}, the gauge dynamics Abelianize to $\mathrm{U}(1)^{N-1}$ at long distances. 
    While these models have taught us much about confinement, they have several features that we do not expect of the dynamics of \emph{non-Abelian} confinement.
    One particularly salient feature common to all of these models is the complete Higgsing of the $\mathrm{S}_N$ subgroup of the $\mathrm{SU}(N)$ gauge group. 
    This Higgsing of $\mathrm{S}_N$ pervades the physics of these theories: it  always gives rise to multiple masses for the dual photons and generically to
    multiple fundamental string tensions \cite{Douglas:1995nw, Argyres:1994xh, Klemm:1994qj} (see Ref.~\cite{Poppitz:2017ivi}
    for a case in which fundamental string tensions remain equal).
    
  The characterization of string tensions is an especially important point of difference between Abelianizing and non-Abelianizing confining gauge theories. Indeed, it is well-known that at asymptotically large distances, the string tensions of confining gauge theories that do not undergo Abelianization
should be solely characterized by center symmetry. This is not the case in the Abelianizing theories mentioned above, at least within the low-energy effective field theory, where string tensions are dictated by charges under $\mathrm{U}(1)^{N-1}$ rather than $N$-ality \cite{Douglas:1995nw, Argyres:1994xh, Klemm:1994qj}. 

Rather curiously, however, it has been observed in numerical experiments that the dynamics of non-Abelian confinement admit an intermediate distance scale where the string tensions are not solely characterized by center symmetry either~\cite{Ambjorn:1984dp, Poulis:1995nn, Bali:2000un, Philipsen:1999wf, Stephenson:1999kh, deForcrand:1999kr, Greensite:2003bk,Wellegehausen:2010ai}: they carry more detailed information on the representations of the gauge group. This naturally suggests that we should try to construct an intermediate theory between Abelian and non-Abelian worlds to leverage what we know about the former to learn more about the latter. As mentioned above, an understanding of unbroken $\mathrm{S}_N$ should be an important clue in this direction.

Thus, the purpose of this work is to construct a 3-dimensional lattice model in which these considerations can be addressed quite explicitly. We call it the \emph{semi-Abelian gauge theory}. To define it, we begin with a pure Abelian lattice model (henceforth to be referred to as the `Abelian model') with gauge group $\mathrm{U}(1)^{N-1}$ such that the permutation group $\mathrm{S}_N$ is present as a global symmetry. The semi-Abelian theory is then obtained by gauging this $\mathrm{S}_N$, and its gauge group is given by
\begin{equation}
    G_{\mathrm{gauge}} = \mathrm{U}(1)^{N-1} \rtimes \mathrm{S}_N. 
\end{equation}
As pure gauge theories are \emph{not} usually equipped with non-Abelian global symmetries, the global or local $\mathrm{S}_N$ symmetry of these models has some rather interesting consequences.

We first show that in both models, the mass gap is generated via Polyakov’s mechanism whereby the proliferation of lattice monopole-instantons results in Debye screening. Crucially, the unbroken (or un-Higgsed) $\mathrm{S}_N$ symmetry implies that the effective potential receives equal contributions from the monopoles associated with the entire $\mathrm{SU}(N)$ root system, which in turn leads to exact degeneracy for the $N-1$ dual photon masses. 
This feature sharply contrasts with the mass generation in the Polyakov model on $\mathbb{R}^3$ or in deformed Yang--Mills on $\mathbb{R}^3 \times S^1$, where the effective potential is sourced at leading order only by the monopoles associated with the (affine) simple roots and the $N-1$ dual photon masses are not degenerate.

After studying these properties of local operators, we move on to study properties of test electric particles, which can be described by the behavior of Wilson loops. In 3 spacetime dimensions, the Coulomb potential is already log-confining, but due to the mass gap generated by the monopole-instantons, the interparticle potential becomes linear-confining. Using the dual formulation of the Wilson loop, we give a semi-classical formula for the string tensions within a reasonable ansatz. We then find that there are infinitely many string tensions. In particular, the semi-Abelian theory furnishes a unique fundamental string tension. 

This, however, raises a puzzle about the string tensions. 
In order to study the spectral properties of the confining forces for test quarks, 
we would like to have some symmetry that acts nontrivially on the Wilson loops. 
One is the well-known center symmetry, which has been recently axiomatized in the framework of higher-form symmetry~\cite{Gaiotto:2014kfa, Sharpe:2015mja}.
Before gauging $\mathrm{S}_N$, our model has a $\mathrm{U}(1)^{N-1}$ $1$-form symmetry, which provides sufficiently strong selection rules to support infinitely many string tensions. 
But after gauging $\mathrm{S}_N$, the center of the gauge group becomes tiny, 
\begin{equation}
    Z(G_{\mathrm{gauge}})=\mathbb{Z}_N,
\end{equation}
so the $1$-form symmetry group becomes $\mathbb{Z}_N$, as it is in $\mathrm{SU}(N)$ Yang--Mills. 
And as we know from $\mathrm{SU}(N)$ Yang--Mills, the $\mathbb{Z}_N$ $1$-form symmetry can only explain the $N$-ality behavior of the string tensions at asymptotically large distances. 
However, the list of string tensions for the semi-Abelian gauge theory turns out to be unchanged by the gauging of $\mathrm{S}_N$. Thus, as in the case of $\mathrm{SU}(N)$ Yang--Mills at intermediate distances, the string tensions of the semi-Abelian theory cannot be dictated by $N$-ality alone. 

We find a resolution to this puzzle in a not-so-obvious but important symmetry of the semi-Abelian  theory, a {\it non-invertible symmetry}. 
Indeed, after the generalization of symmetry to higher-form symmetry, it has been recognized that the most essential feature of a conservation law is the existence of topological defects, at least in the context of relativistic quantum field theories (QFTs).  
In other words, as long as one keeps intact the existence of topological defect operators, one may make up a new kind of symmetry by replacing or weakening other features of the generalized global symmetry of Ref.~\cite{Gaiotto:2014kfa} (e.g. higher-group symmetry~\cite{Kapustin:2013uxa, Sharpe:2015mja, Cordova:2018cvg, Benini:2018reh, Misumi:2019dwq,  Tanizaki:2019rbk, Hidaka:2020iaz, Hidaka:2020izy}).  
Non-invertible symmetries are also generated by topological defects, but their fusion rules do not conform to the usual group multiplication; as the name suggests, a non-invertible symmetry transformation need not have an inverse (which of course never occurs if the transformations form a group).
The notion of non-invertible symmetry is still in its infancy, and it seems that its mathematical formulation has been so far established only in $2$-dimensional spacetimes.
Nevertheless, the utility of such topological operators in probing quantum systems has been elucidated in several recent studies, as the notion of symmetry itself tends to be broadened ~\cite{Bhardwaj:2017xup, Buican:2017rxc, Freed:2018cec,  Chang:2018iay,Thorngren:2019iar,Ji:2019jhk, Rudelius:2020orz, Komargodski:2020mxz, Aasen:2020jwb}.
In that context, the new symmetry goes by various names, such as non-invertible symmetry, categorical symmetry, etc.
Here, we would like to emphasize that the non-invertible symmetry clarifies an important feature of our $3$-dimensional semi-Abelian gauge theory. 
Thanks to the simplicity of the model, the symmetry considerations we propose can be checked against concrete calculation. 

We construct a generator of a continuous non-invertible symmetry, and compute its action on several Wilson loops. 
By looking at its eigenvalues, we show that we can distinguish different string tensions even if they correspond to representations of
the same $N$-ality. 
We also discuss conditions where such extra selection rules by noninvertible symmetry are lost by the addition of dynamical electric particles, and we compare them with the standard string-breaking arguments to check that they are consistent. 
Finally, as an application, we discuss an example where the non-invertible symmetry is explicitly broken to a discrete sub-symmetry, so that even though the number of string tensions becomes
finite, there still remain some string tensions beyond $N$-ality.

\section{3d $\mathrm{U}(1)^{N-1}$ lattice gauge theory with $\mathrm{S}_N$ global symmetry}
\label{sec:Abelian_lattice}

There are two basic models  that we study in this paper:
\begin{itemize} 
\item $\text{U}(1)^{N-1}$  Abelian gauge theory with discrete non-Abelian global symmetry ${\rm S}_N$
\item    $\text{U}(1)^{N-1} \rtimes \mathrm{S}_N$  semi-Abelian gauge  theory   
\end{itemize}
The second one can be obtained  by gauging  the $\mathrm{S}_N$ 0-form symmetry of the first one, and the main purpose of this paper is to understand its properties. 
To this end, we must first understand the properties of the first theory, and this is the goal of this section.

\subsection{Description of the $\mathrm{U}(1)^{N-1}$ lattice gauge theory} 
\label{SecDescription}

The $\text{U}(1)^{N-1}$ lattice gauge model with the $\mathrm{S}_N$ global symmetry can be realized either by a standard Wilson-type formulation~\cite{Wilson:1974sk}  or by a  Villain-type formulation~\cite{Villain:1974ir}. Since these provide somewhat complementary perspectives, we end up working with both.

\subsubsection{Villain formulation}
To give the Villain formulation, we consider a link field $\bm{A}_\ell$ valued in $\mathbb{R}^{N-1}$ and a plaquette field $\bm{n}_p$ valued in the root lattice $\G_{\rm r} \subset \mathbb{R}^{N-1}$ of SU($N$).
We take for the action 
\begin{equation}
    S = \frac{1}{4 \pi e^2} \sum_p (\bm{F}_p + 2\pi \bm{n}_p)^2
    \label{multiVillain}
\end{equation}
where $\bm F_p = (\dd \bm A)_p$ is the field-strength and $e$ is the gauge coupling. The partition function is given by
\begin{equation}
    Z = \sum_{\{\bm{n}_p \in \G_{\rm r} \}} \int_{\mathbb{R}^{N-1}}[ \dd \bm{A}_\ell ]\,  \ee^{-S} \, .
    \label{multiVillainPartition}
\end{equation}
This theory is invariant under the $0$-form gauge symmetry
\begin{equation}
    \bm{A}_\ell \rightarrow \bm{A}_\ell + (\dd \bm{\lambda})_\ell \, , 
    \qquad 
    \bm{\lambda}_s \in \mathbb{R}^{N-1}, 
    \label{0-form-gauge}
\end{equation}
and the $1$-form gauge symmetry
\begin{equation}
    \bm{A}_\ell \rightarrow \bm{A}_\ell + 2\pi \bm{\beta}_\ell \,,
    \qquad \bm{n}_p \rightarrow \bm{n}_p - (\dd \bm{\beta})_p \,,
    \qquad \bm{\beta}_\ell \in \G_{\rm r} \,. 
    \label{1-form-gauge}
\end{equation}
In view of the fact that $\mathbb R^{N-1} /(2\pi \G_{\rm r}) \simeq {\rm U}(1)^{N-1}$, we see that this indeed defines a ${\rm U}(1)^{N-1}$ gauge model. 

One way to understand this Villain-type formulation  \cite{Polyakov:1975rs, Sulejmanpasic:2019ytl}  is to imagine that we had begun with pure $\mathbb R^{N-1}$ gauge theory,
\begin{equation}
    Z = \int_{\mathbb{R}^{N-1}} [\dd \bm A_\ell] \exp \left( - \frac{1}{4 \pi e^2} \sum_p \bm F_p^2 \right),
\end{equation}
and then considered gauging the discrete subgroup $2\pi\G_{\rm r}$ of the $\mathbb R^{N-1}$ 1-form center symmetry group, which acts according to
\begin{equation}
    \bm A_\ell \mapsto \bm A_\ell + \bm \theta_\ell \,, \qquad \bm \theta_\l \in \mathbb R^{N-1} \,, \qquad (\dd \bm \theta)_p = 0 \,.
    \label{1-form-global}
\end{equation}
The simplest way to do that is to introduce the discrete $\G_{\rm r}$-valued plaquette field $\bm{n}_p$, and then demand that the local transformations \eqref{1-form-gauge} be gauge redundancies. Minimal coupling to the field $\bm{n}_p$ would then produce the action \eqref{multiVillain} and the partition function \eqref{multiVillainPartition}.

\paragraph{Global symmetries:}
Let us now discuss the global symmetries of this model. First, as already noted above, there is a ${\rm U}(1)^{N-1}$ 1-form center symmetry \eqref{1-form-global}, where the group is ${\rm U}(1)^{N-1}$ rather than $\mathbb R^{N-1}$ thanks to the 1-form gauge structure \eqref{1-form-gauge}. 

Importantly, the theory has   a discrete  non-Abelian 0-form global symmetry, 
\begin{equation}
    \bm{A}_\ell \mapsto \Pi \bm{A}_\ell \, ,
    \qquad 
    \bm{n}_p \mapsto \Pi \bm{n}_p, 
    \label{0-form-global}
\end{equation}
under O($N-1$) transformations $\Pi$ that preserve the root lattice $\G_{\rm r}$. Such transformations constitute the automorphism group of the SU($N$) root system, and therefore the symmetry group here is 
\begin{equation}
G^{[0]}_{\mathrm{global}} = \left\{
\begin{matrix}
\mathrm{S}_N \rtimes \mathbb{Z}_2 & \quad (N>2),\\
\mathrm{S}_2\simeq \mathbb{Z}_2 & \quad (N=2).
\end{matrix}\right.  
\label{ngs}
\end{equation}

 The $\mathrm{S}_N$ corresponds to the Weyl group of SU($N$), which is generated by the reflections in the hyperplanes orthogonal to the roots
\begin{equation}
    \bm{A}_\ell \mapsto \bm{A}_\ell - \bm{\alpha} (\bm{\alpha} \cdot \bm{A}_\ell) \,,
    \qquad \bm{n}_p \mapsto \bm{n}_p - \bm{\alpha} (\bm{\alpha} \cdot \bm{n}_p) \,, 
    \qquad \bm{\alpha} \in \Phi,
    \label{Weyl}
\end{equation}
where $\Phi$ is the set of roots for SU($N$). The pair $(\bm A_\l, \bm {n}_p)$ thus transforms in the standard representation $D_{\rm std}$ of $\mathrm{S}_N$, which is the $(N-1)$-dimensional irreducible representation. Meanwhile, the $\mathbb{Z}_2$ is simply generated by the reflection
\begin{equation}
    \bm{A}_\ell \mapsto - \bm{A}_\ell \,, 
    \qquad \bm{n}_p \mapsto - \bm{n}_p, 
\end{equation}
which we may think of as charge conjugation. We note that, for $N=2$, these two operations are identical. 

Note that the existence of the non-Abelian global symmetry \eqref{ngs} is somewhat  unusual for a pure gauge theory.   In general,  pure gauge theories without matter fields, either Abelian or non-Abelian,  do not possess  {\it non-Abelian} global symmetries.  In the  ${\rm U}(1)^{N-1}$ gauge theory we are considering,  this symmetry is present. The gauging  of the permutation group $\mathrm{S}_N$ will generate a genuinely non-Abelian gauge theory, which we shall investigate.

The basic observables we are concerned with are the Wilson loops, which are here given by
\begin{equation}
    W_{\bm w}(C) = \exp \left( \ii \int_C \bm{w} \cdot \bm{A} \right),
\end{equation}
with $\bm{w}$ in the weight lattice $\G_{\rm w}$ of ${\rm{SU}}(N)$. Note that it is invariance under the 1-form gauge transformations \eqref{1-form-gauge} that requires the electric charge to be a weight. The Wilson lines transform under the 0-form discrete symmetry \eqref{0-form-global} as
\begin{equation}
    W_{\bm w}(C) \mapsto W_{\Pi^{-1} \bm{w}}(C),
\end{equation}
and under the 1-form center symmetry \eqref{1-form-global} as
\begin{equation}
    W_{\bm w} (C) \mapsto W_{\bm w} (C) \exp \left( \ii \int_C \bm{w} \cdot \bm \theta \right). 
\end{equation}

\subsubsection{Wilson formulation}

To construct the $\text{U}(1)^{N-1}$  lattice gauge theory in the Wilson formulation, we consider 
$N$  gauge fields $a^1_\ell, \ldots ,a^N_\l$     and  a Lagrange multiplier   $v_\l$  which  is an integer-valued link-field. 
The  dynamics is determined by the action
\begin{equation}
    S_{\rm W} = \b \sum_p \sum_{i=1}^N (1 - \cos f^i_p) - \ii \sum_\l \sum_{i=1}^N v_\l a^i_\l , 
	\label{eq:Wilson_Formulation}
\end{equation}
where the $f^i_p = (\dd a^i)_p$ are the field-strengths.  
In the partition function, we integrate over $a^i_\ell \in [0,2\pi]$ and sum over $v_\l \in \mathbb Z$:
\begin{equation}
    Z = \sum_{\{ v_\l \in \mathbb{Z} \}} \int_0^{2\pi} [ \dd a^i_\ell ] \ee^{-S_{\mathrm W}}. 
    \label{multiWilsonPartition}
\end{equation}
In particular, summation over $v_\l$ in the partition function produces the constraint
\begin{equation}
    \sum_{i=1}^N a^i_\l = 0 \mod 2 \pi,
\end{equation}
so that only $N-1$ of the photons are physical.\footnote{
We could integrate out $v_\ell$ and any one of the photon fields. Then after some simple field redefinitions, we would obtain the action
\begin{equation}
    S_{\mathrm W} = \beta \sum_p \sum_{i=1}^N (1 - \cos ( {\bm \n}_i \cdot {\bm f}_p)),
\end{equation}
where the ${\bm \n}_i$ are the weights of the defining representation of SU($N$) and ${\bm f}_p$ is the field-strength of an ($N-1$)-component Abelian gauge field ${\bm a}_\ell$.
}

One nice thing about this formulation is that the $\mathrm{S}_N$ symmetry is manifest; it acts simply by permuting the $N$ photons:
\begin{equation}
    (a^1_\l , \ldots , a^N_\l) \mapsto (a^{P(1)}_\l , \ldots , a^{P(N)}_\l) \,, \qquad P\in \mathrm{S}_N. 
\end{equation}

For now, we shall prefer to work with the Villain form over the Wilson one, because the former enjoys exact dualities which allow us to analyze the dynamics most simply. Nevertheless, the two formulations are equivalent at weak coupling, as we demonstrate in Appendix~\ref{SecWilsonFormulation}. Later on, in Section~\ref{sec:SemiAbelian} where we gauge the $\mathrm{S}_N$ global symmetry, we will find the Wilson form more convenient.

\subsection{Mass gap and spectrum}
\label{sec:mass_gap_and_spectrum}

In this subsection, we discuss the mass gap of the lattice Abelian gauge theory with $\mathrm{S}_N$ global symmetry. 
  
First, as we shall review in Section~\ref{SecDualFormulations}, we note that the Villain form is {\it exactly} dual to a multi-component Coulomb gas; that is, the partition function can rewritten in the form
\begin{equation}
    Z= \sum_{\{\bm{q}(\xt) \in \G_{\text{r}} \}} \exp \left( - \frac{\pi}{e^2} \sum_{\xt,\xt'} v(\xt -\xt') \bm{q}(\xt) \cdot \bm{q} (\xt') \right) ,
    \label{MultiCoulomb}
\end{equation}
where $v(\xt)$ is the lattice Coulomb potential, and $\bm q (\xt)$ is a $\G_{\rm r}$-valued scalar field on the dual lattice. 
Here, one can interpret $\{\bm q({\xt})\}$ as a configuration of magnetic monopoles; $\bm q(\xt)$ is the magnetic charge of the magnetic monopole at $\xt$. 
As is familiar, the proliferation of monopoles in the Euclidean description of the vacuum results in Debye screening, and hence, the correlation length remains finite for any nonzero value of the coupling \cite{Polyakov:1975rs}. While this is more or less self-evident, we can go further and obtain the long-distance effective field theory: 
\begin{equation}
    Z = \int \Diff \bm{\sigma} \exp \left( - \frac{e^2}{2 \pi} \int \dd^3 x \, \left\{ \frac{1}{2} | \dd \bm{\sigma}|^2 + M^2 \sum_{\bm{\alpha} \in \Phi^+} \bigl(1-\cos( \bm{\alpha} \cdot \bm{\sigma})\bigr) \right\} \right),
    \label{multiEFT}
\end{equation}
where the dual photon field $\bm \s$ is a $2 \pi \G_{\mathrm{w}}$-periodic scalar, $\Phi^+$ is a set of positive roots for SU($N$), and $M^2 \propto \ee^{-\text{const.}/e^2}/e^2$. 
This effective description shows very clearly the presence of a nonzero mass gap. It will be derived in Section~\ref{long-distance}.

We can immediately observe that the $N-1$ dual photons must have exactly the same mass. 
The degeneracy is a consequence of the $\mathrm{S}_N$ global symmetry inherited from the microscopic theory. 
To see this, note that the dual photons $\bm \sigma$ transform in the standard representation $D_{\rm std}$ of $\mathrm{S}_N$:
\begin{equation}
     \bm\s \mapsto D_{\rm std}(P) \bm \s \,,
    \qquad P \in \mathrm{S}_N. 
\end{equation}
The mass matrix for the dual photons,
\begin{equation}
    (M^2_{\bm{\sigma}})^{ij} = M^2 \sum_{\bm \a \in \Phi^+} \a^i \a^j,
\end{equation}
is also invariant under the $\mathrm{S}_N$ transformation,
\begin{equation}
   D_{\rm std}(P) M^2_{\bm{\sigma}} D_{\rm std}^{-1}(P) = M^2_{\bm{\sigma}}, \qquad P \in \mathrm{S}_N.
\end{equation}
Since $D_{\rm std}$ is irreducible, it follows from Schur's lemma that $M^2_{\bm{\sigma}}$ must be proportional to the identity matrix. 
The mass gap is thus the $(N-1)$-fold degenerate eigenvalue of $M_{\bm \s}^2$. By taking the trace of $M_{\bm \s}^2$ and using $\bm \a^2 = 2$, one easily finds the mass gap to be 
\begin{equation}
    M_{\mathrm{gap}} = \sqrt{N}M\propto \frac{\sqrt{N}}{e} \ee^{-\text{const.}/e^2}.
\end{equation}

\subsubsection{Multi-component Coulomb gas representation of the Villain form}

\label{SecDualFormulations}

Here we show that the Villain form \eqref{multiVillainPartition} of our theory is exactly dual to multi-component Coulomb gas \eqref{MultiCoulomb}, using standard techniques in Abelian lattice gauge theory \cite{Polyakov:1987ez, Banks:1977cc, Savit:1977fw, Gopfert:1981er}. 
We derive the equivalence very briefly here, but the detailed derivation for the single-component $\rm{U}(1)$ gauge theory is reviewed in Appendix~\ref{SecAbelianDuality}.

We first note that the Poisson summation formula can be generalized on the weight and root lattices to give 
\begin{equation}
    \sum_{\bm{n}_p\in \Gamma_{\mathrm{r}} } \exp\left(- \frac{1}{4 \pi e^2}(\bm{F}_p+2\pi \bm{n}_p)^2\right)
    = \sum_{\bm{k}_p\in \Gamma_{\mathrm{w}}} \exp\left(- \pi e^2 \bm{k}_p^2+\im \bm{k}_p\cdot \bm{F}_p\right)
    \label{eq:Poisson_general}
\end{equation}
up to an overall coefficient. By performing the ${\bm A}_\ell$ integration exactly, we obtain the constraint
$
    (\diff^\dagger \bm{k})_\ell=0, 
$
which can be easily solved by setting 
\begin{equation}
    *\bm{k}=\diff \bm{m}, 
    \label{eq:duality_k_m}
\end{equation}
where $\bm{m} (\tilde{x})$ is a $\Gamma_{\mathrm{w}}$-valued scalar field on the dual lattice.\footnote{
For clarity, we ignore the effect of nontrivial spacetime topology.}
After this replacement, the partition function becomes
\begin{equation}
    Z=\sum_{\{\bm{m} (\tilde{x}) \in \Gamma_{\mathrm{w}}\}} \exp\left(- \pi e^2 \sum_{\lt} (\diff \bm{m})_{\lt}^2\right).
    \footnote{This representation may be thought of as a `$\G_{\mathrm w}$-ferromagnet' by analogy with the corresponding expression with $\mathbb{Z}$ in place of $\G_{\mathrm w}$.  The $\G_{\mathrm w}$-ferromagnet representation is exactly dual to the $\G_{\rm r}$-component Coulomb gas representation \eqref{MultiCoulomb}.  While the latter converges rapidly at weak coupling $e^2 \rightarrow 0$, the former converges rapidly at strong coupling $e^2 \rightarrow \infty$.  }
\end{equation}
We now wish to replace $\bm{m}(\xt)$ by a continuous field; it can be done with the help of the Poisson summation formula again, this time in the form
\begin{equation}
    \sum_{\bm{m}(\xt)\in \Gamma_{\mathrm{w}}}\delta(\bm{\sigma}(\xt)-2\pi \bm{m}(\xt))=\sum_{\bm{q}(\xt)\in \Gamma_{\mathrm{r}}} \exp(\im\, \bm{q}(\xt)\cdot \bm{\sigma}(\xt)), 
\end{equation}
which introduces the dual photon field $\bm \s (\xt)$. The result is 
\begin{equation}
    Z=\int [\dd \bm{\sigma} (\xt) ] \sum_{\{\bm{q}(\xt)\in \Gamma_{\mathrm{r}}\}} \exp\left(-\frac{e^2}{4 \pi}\sum_{\xt} (\partial_\mu^- \bm{\sigma}(\xt))^2 + \im \sum_{\xt} \bm{q}(\xt)\cdot\bm{\sigma}(\xt)\right). 
\end{equation}
After performing the Gaussian integration over $\bm{\sigma}$, we arrive at the multi-component Coulomb gas representation (\ref{MultiCoulomb}).

\subsubsection{Long-distance effective theory}
\label{long-distance}
We now wish to pass to the long-distance effective description \eqref{multiEFT} \cite{Polyakov:1987ez, Banks:1977cc, Savit:1977fw, Gopfert:1981er}. To this end, we first split the Green function $\D^{-1}$ in (\ref{MultiCoulomb}) into two parts by adding and subtracting $(\Delta +M_{\mathrm{PV}}^2)^{-1}$: 
\begin{equation}
    \D^{-1} 
    = \D^{-1} (1 + \D/M_{\mathrm{PV}}^2)^{-1}  +  (\D +M_{\mathrm{PV}}^2)^{-1} = u_{M_{\rm PV}} + w_{M_{\rm PV}}.
    \label{PV}
\end{equation}
Here, $u_{M_{\mathrm{PV}}}(\xt)$ is the Green function of the Pauli--Villars regulated Laplacian $\Delta_{M_{\mathrm{PV}}}\equiv \D (1 + \D/M_{\mathrm{PV}}^2)$, and  $w_{M_{\mathrm{PV}}}(\xt)$ is the Yukawa Green function. 
Since $w_{M_{\mathrm{PV}}}(\xt)$ decays exponentially fast, we can take $w_{M_{\mathrm{PV}}}(\xt)  = w_{M_{\mathrm{PV}}}(0) \delta_{\xt , 0}$. Furthermore, it is straightforward to show that $w_{M_{\mathrm{PV}}}(0) = v(0) - {\cal O}(1/M_{\mathrm{PV}}) \approx  0.253 - {\cal O}(1/M_{\mathrm{PV}})$ \cite{Ukawa:1979yv}. 

With the above decomposition, we rewrite the Coulomb gas partition function as
\begin{equation}
    Z = \sum_{\{\bm{q}(\xt) \in \G_{\text{r}} \}} 
    \exp \left( - \frac{\pi}{e^2} \sum_{\xt,\xt'} u_{M_{\mathrm{PV}}}(\xt -\xt') \bm{q}(\xt) \cdot \bm{q} (\xt') 
    -\frac{1}{2}I\sum_{\xt} \bm q(\xt)^2\right) ,
    \label{MultiCoulombPV}
\end{equation}
where $I \equiv 2 \pi v(0)/e^2$, and then reintroduce the dual photon field $\bm \s $ to get
\begin{equation}
    Z = \int [\dd \bm \s (\xt)] \ee^{-\frac{e^2}{4 \pi }\sum_{\xt}\bm \s(\xt) \Delta_{M_{\mathrm{PV}}} \bm \s(\xt)} \sum_{\{\bm q(\xt) \in \G_{\text{r}} \}} \ee^{\ii \sum_{\xt} \bm q(\xt) \bm \s(\xt)} \ee^{-\frac{1}{2}I\sum_{\xt} \bm q(\xt)^2}. 
    \label{ExactDualPhoton}
\end{equation}

At this point, we want to perform a cluster expansion of the partition function. For weak coupling, $I$ is large, and so $\ee^{-I}$ is exponentially small. Thus, at leading order in semi-classics, we can restrict the summation over $\bm q(\xt)$ to $\{0\} \cup \Phi$. Indeed, ${\bm \a}^2 = 2$ for each $ \bm \a \in \Phi$, so all the monopoles whose charges are roots have the same minimal action $I$, and there are $N(N-1)$ degenerate saddles at leading order in semi-classics. 
Performing the summation over ${\bm q}(\xt)$ with this restriction then yields 
\begin{equation}
    \sum_{\bm q(\xt) \in \{0\} \cup \Phi} \ee^{\ii \bm q (\xt) \bm \s (\xt)} \ee^{-\frac{1}{2}I \bm q(\xt)^2}
    \approx \exp \left( 2 \ee^{- I} \sum_{\bm \a \in \Phi^+} \cos( \bm \a \cdot\bm \s (\xt)) \right). 
\end{equation}
Finally, inserting this into \eqref{ExactDualPhoton}, we get
\begin{equation}
    Z = \int [\dd \bm \s (\xt)] \exp \left(-\frac{e^2}{4 \pi}\sum_{\xt}\bm \s(\xt) \Delta_{M_{\mathrm{PV}}} \bm \s(\xt) + 2 \ee^{- I} \sum_{\xt} \sum_{\bm {\a} \in \Phi^+} \cos( \bm \a \cdot \bm \s (\xt) ) \right),
    \label{LatticeMultiEFT}
\end{equation}
which, upon taking the continuum limit, coincides with \eqref{multiEFT}.  

We note that, in this derivation, we have neglected the effect of the spacetime topology, and thus the periodicity of the dual photon field $\bm \sigma$ is undetermined. 
Had we taken it into account, we would have identified $\bm \sigma$ as a $2\pi\Gamma_\mathrm{w}$-periodic scalar. (We elaborate on this subtlety in Appendix~\ref{SecAbelianDuality}.)

\paragraph{Remarks:} The fact that the sum over monopoles goes over all roots ${\bm \a} \in \Phi$ and that all monopoles associated with these roots have the same action distinguishes our ${\rm U}(1)^{N-1}$ lattice gauge theory with $\mathrm{S}_N$ symmetry from Yang--Mills adjoint Higgs systems which exhibit  dynamical Abelianization $\mathrm{SU}(N) \rightarrow  {\rm U}(1)^{N-1}$.\footnote{
It is also worth nothing that in $\mathcal{N}=4$ $\mathrm{SU}(N)$ super Yang--Mills theory softly broken down to  $\mathcal{N}=1^*$ on $\mathbb{R}^3\times S^1$ as well, it is necessary to sum over monopoles associated with non-simple roots in order to capture the ground state properties correctly~\cite{Dorey:1999sj}. This data is encoded in an elliptic superpotential, but the $\mathrm{S}_N$  symmetry is still Higgsed in  generic vacua.}
In the latter, if the adjoint Higgs are algebra-valued, as in the Polyakov model~\cite{Polyakov:1975rs}, 
the sum over monopoles at leading order in semi-classics is restricted to the  $N-1$ simple roots ${\bm \a} \in \Delta$, while if the adjoint Higgs are group-valued, as in deformed Yang--Mills~\cite{Unsal:2008ch}, the sum over monopoles is restricted to the $N$ affine simple roots.  
There are monopoles associated with non-simple roots as well, but these are higher action and do not contribute at leading order; in general, the monopoles split into $\mathbb{Z}_N$-orbits with hierarchical fugacities  $ {\ee}^{-S_0} \gg  {\ee}^{-2 S_0} \gg \cdots \gg {\ee}^{-(N-1) S_0}$. 
In our present construction, $\mathrm{S}_N$ permutation symmetry guarantees that all $N(N-1)$ monopoles associated with the roots have the same action. 
In theories like the Polyakov model and Seiberg--Witten theory, $\mathrm{S}_N$ is part of the gauge structure of the microscopic theory, but it is spontaneously broken by the vacuum expectation value of the Higgs field which imposes an ordering on the eigenvalues of the adjoint Higgs.
These models therefore exhibit $\mathcal{O}(N)$ different types of fundamental string tensions. 
We will see how the string tensions behave in our $\mathrm{U}(1)^{N-1}$ Abelian model in the following subsection.

\subsection{Wilson loops and string tensions}\label{sec:string_tension_Abelian}

In this subsection, we show that the Abelian gauge model confines and we approximately determine the string tensions. 

\begin{figure}
\vspace{-1cm}
\begin{center}
\includegraphics[width = 1.1\textwidth]{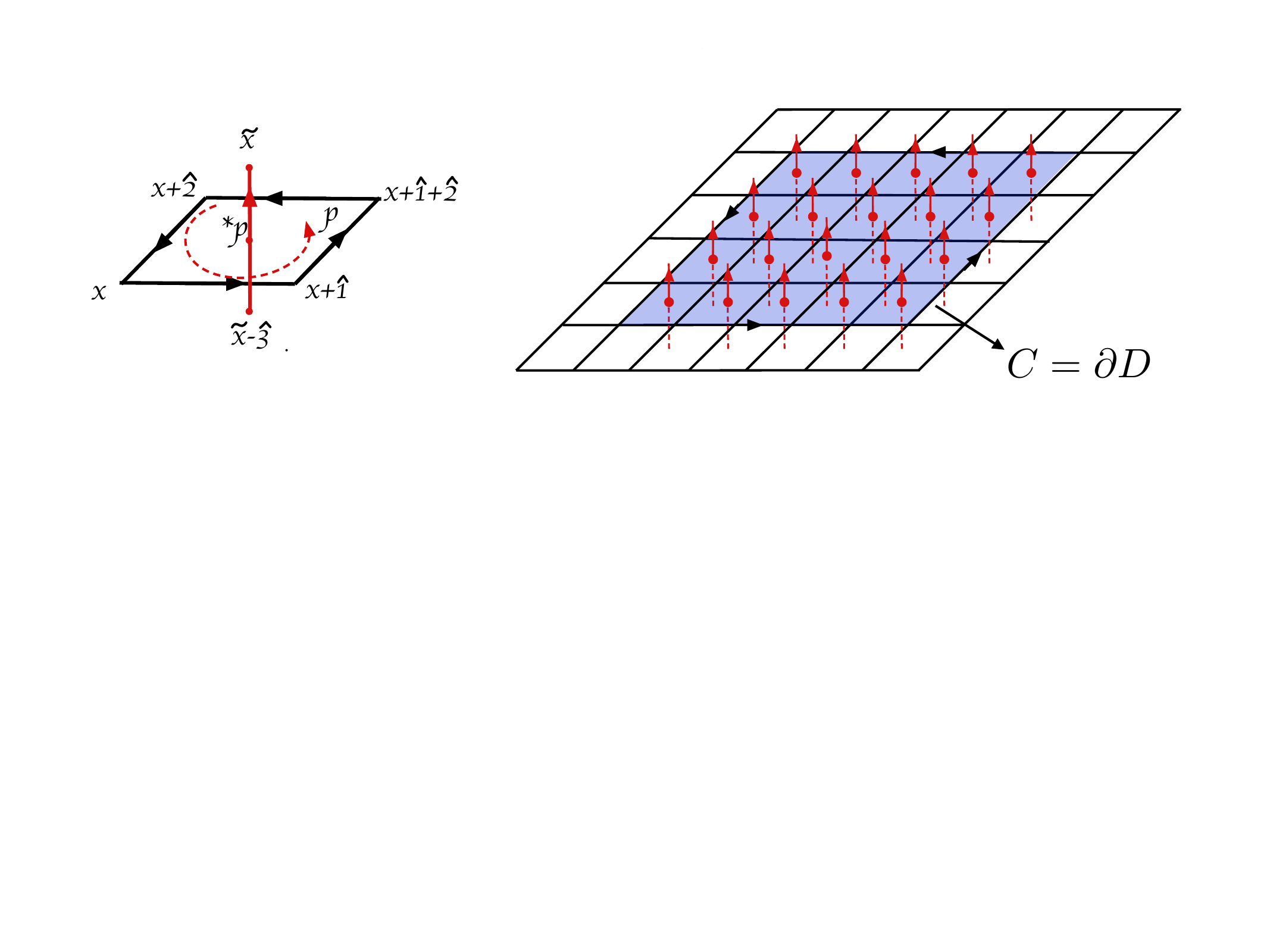}
\vspace{-7.5cm}
\caption{Left: The dual of the plaquette $p$ is the link $*p$ on the dual lattice intersecting $p$ as shown. 
Right: $D$ is the shaded region bounded by the curve $C$. The Poincare dual $[D]$ is a bump 1-form on the dual lattice that is 1 on each of the red links shown, and 0 everywhere else.    }
\label{fig:poincare}
\end{center} 
\vspace{-0.5cm}
\end{figure}

We begin by showing how a Wilson loop $W_{\bm{w}}(C)=\exp\left(\im \int_C \bm{w}\cdot \bm{A}\right)$ with electric charge ${\bm w} \in {\G}_{\rm w}$
is computed in the long-distance effective theory \cite{Polyakov:1987ez}. 
For our purposes, it will suffice to take $C$ to be a contractible loop, so that it is the boundary of a 2-dimensional surface $D$. We can then write
\begin{equation}
     W_{\bm{w}}(C)= \exp\left(\im \int_D \bm{w}\cdot \bm{F}\right) = \exp\left(\im \sum_p [D]_{*p} (\bm{w}\cdot\bm{F}_p)\right). 
\end{equation}
Here we have introduced the Poincar\'e dual $[D]$ of $D$; it is a bump $1$-form on the dual lattice (see  
Figure~\ref{fig:poincare})
given by 
\begin{equation}
    [D]_{*p} = 
    \begin{cases}
    1, \qquad \text{if} \, \, \, p \subset D, \\
    0, \qquad \text{otherwise}.
    \end{cases}
\end{equation}

Now let us repeat the derivation of the dual theory, this time with the insertion of the Wilson loop. Using the Poisson resummation (\ref{eq:Poisson_general}), the path-integral weight becomes 
\begin{equation}
    \exp\left(- \pi e^2 \bm{k}_p^2+\im \bm{k}_p\cdot \bm{F}_p+\im [D]_{*p} (\bm{w}\cdot \bm{F}_p)\right), 
\end{equation}
where the last term comes from the Wilson loop. The integration over $\bm{A}$ produces the constraint, 
$\diff^\dagger(\bm{k}+*(\bm{w}[D]))=0$, 
which can be solved by
\begin{equation}
    *\bm{k}=-\bm{w}[D]+\diff \bm{m}, 
\end{equation}
instead of (\ref{eq:duality_k_m}). 
At this point, the rest of the derivation proceeds exactly as before, and we obtain 
\begin{equation}
    \langle W_{\bm{w}}(C)\rangle = \int \Diff \bm{\sigma} \exp \left( - \frac{e^2}{2 \pi} \int \dd^3 x \, \left\{ \frac{1}{2} \Bigl|\diff \bm{\sigma}-2\pi\bm{w}[D]\Bigr|^2 + M^2 \sum_{\bm{\alpha} \in \Phi^+} \bigl(1-\cos( \bm{\alpha} \cdot \bm{\sigma})\bigr) \right\} \right).
    \label{WilsonLoopDual}
\end{equation}
This expression shows that the Wilson loop is realized as a defect operator in the dual formulation. 
That is to say, it is evaluated by removing the loop $C$ from the spacetime and restricting the path integral to configurations satisfying $\oint_{S^1}\diff \bm{\sigma}=2\pi \bm{w}$ for small loops $S^1$ that link with the loop $C$. 
Taking the ratio with the unconstrained path integral, we obtain the expectation value of the Wilson loop. 

We are now in a position to approximately determine the string tensions. It will suffice to compute the functional integral in \eqref{WilsonLoopDual} in the classical approximation. For convenience, let us take the loop $C$ as well as $D$ to lie in the $z \equiv x_3 = 0$ plane.
If we take the loop $C$ to be so large that $D$ essentially fills the $z = 0$ plane, then the action density localizes around $z=0$. 
As a result, the area law decay for the Wilson loop $W_{\bm w}(C)$ is observed,
\begin{equation}
    \langle W_{\bm{w}}(C)\rangle \sim \exp\Bigl(-T_{\bm{w}}\, \mathrm{Area}(D)\Bigr),
\end{equation}
and its string tension is given by the minimal action density
\begin{equation}
    T_{\bm{w}} 
    = \min_{\bm{\sigma}(z)}\, \frac{e^2}{2 \pi} \int_{-\infty}^{+ \infty} \dd z \, \left\{ \frac{1}{2} \left(\frac{\dd \bm \sigma}{\dd z} \right)^2 + M^2\sum_{\bm{\alpha} \in \Phi^+}\Bigl(1-  \cos( \bm{\alpha} \cdot \bm \sigma)\Bigr) \right\},
    \label{Tension}
\end{equation}
with the boundary condition 
\begin{equation}
    \bm{\sigma}(-\infty)=0,\qquad \bm{\sigma}(+\infty)=2\pi\bm{w}. 
\end{equation}
To go further, let us take as a plausible ansatz
\begin{equation}
    \bm{\sigma}(x)=\bm{w}\sigma(z),
    \label{eq:ansatz_string}
\end{equation}
and then proceed to evaluate the string tension analytically within this ansatz. Substituting (\ref{eq:ansatz_string}) into (\ref{Tension}), we obtain 
\begin{eqnarray}
T_{\bm{w}}&=&
\min_{{\sigma}(z)}\, \frac{e^2}{2 \pi} \int_{-\infty}^{+ \infty} \dd z \, \left\{ \frac{\bm{w}^2}{2} \left(\frac{\dd \sigma}{\dd z} \right)^2 + M^2\sum_{\bm{\alpha} \in \Phi^+}\Bigl(1-  \cos( \bm{\alpha} \cdot \bm{w} \sigma)\Bigr) \right\}\nonumber\\
&=&\frac{e^2M}{\pi} \int_{0}^{2\pi} \dd \sigma \, \sqrt{ \frac{\bm{w}^2}{2}  \sum_{\bm{\alpha} \in \Phi^+}\Bigl(1-  \cos( \bm{\alpha} \cdot \bm{w} \sigma)\Bigr) }, 
\label{eq:BPS_string_tension}
\end{eqnarray}
which is the Bogomol'nyi–Prasad–Sommerfield (BPS) bound~\cite{Bogomolny:1975de, Prasad:1975kr}. 
Although this is just an upper bound for the actual string tension, we assume that it gives a reasonable estimate.

\subsubsection{Explicit evaluation of string tensions}

Using the formula (\ref{eq:BPS_string_tension}), we shall evaluate the string tensions explicitly for a few cases. 
Here, we take $\bm{w}=\bm{\mu}_1,\, 2\bm{\mu}_1,\, \bm{\mu}_2$, which correspond to the highest weights of the fundamental, symmetric, and anti-symmetric representations of $\mathrm{SU}(N)$, respectively. 
We will also comment on the case $\bm{w}=\bm{\alpha}\in \Gamma_{\mathrm{r}}$, corresponding to the adjoint representation of $\mathrm{SU}(N)$. 

Let us start with $\bm{w}=\bm{\mu_1}$, which is the highest weight of the fundamental representation of $\mathrm{SU}(N)$. We obtain \begin{eqnarray}
T_{\bm{\mu}_1}&=&{e^2M\over \pi}{N-1\over \sqrt{2N}}\int_0^{2\pi}\diff \sigma \sqrt{1-\cos(\sigma)}\nonumber\\
&=&{4e^2 M \over \pi}{N-1\over \sqrt{N}}. 
\end{eqnarray}
We will use this quantity as a unit for the other string tensions. 

We next consider $\bm{w}=2\bm{\mu}_1$, the highest weight of the $\mathrm{SU}(N)$ two-index symmetric representation. Since $\sigma$ wraps $S^1$ twice, we find that 
\begin{equation}
    T_{2\bm{\mu}_1}=2T_{\bm{\mu}_1}. 
\end{equation}
Thus, the symmetric string tension is twice the fundamental one, which suggests that the symmetric string can be interpreted as the sum of two independent fundamental strings.  
The multi-string ansatz is also a candidate, which may give a reasonable approximation of confining strings~\cite{Anber:2015kea}, so we will compare it with (\ref{eq:ansatz_string}) for other strings, too. 

For the two-index anti-symmetric string, $\bm{w}=\bm{\mu}_2$, we find 
\begin{eqnarray}
T_{\bm{\mu}_2}&=&
{e^2M\over \pi}{\sqrt{2}(N-2)\over \sqrt{N}} \int_0^{2\pi}\diff \sigma \sqrt{1-\cos(\sigma)}\nonumber\\
&=&{8e^2M \over \pi}{N-2\over \sqrt{N}}\nonumber\\
&=&{2(N-2)\over N-1}T_{\bm{\mu_1}}.
\end{eqnarray}
For $N=3$, $\bm{\mu}_2$ gives the conjugate representation of $\bm{\mu}_1$, and in this case we indeed find that $T_{\bm{\mu}_2}=T_{\bm{\mu}_1}$. For $N>3$, we find $T_{\bm{\mu}_1}< T_{\bm{\mu}_2}<T_{2\bm{\mu}_1}=2T_{\bm{\mu_1}}$, and $T_{\bm{\mathrm{\mu}_2}} \approx 2T_{\bm{\mu}_1}$ for $N\gg 1$. 
Since $\bm{\mu}_2=\bm{\mu}_1+(\bm{\mu}_1-\bm{\alpha}_1)$, we can understand this upper bound $2T_{\bm{\mu}_1}$ as a sum of two-independent fundamental strings $2T_{\bm{\mu}_1} =T_{\bm{\mu}_1}+T_{\bm{\mu}_1-\bm{\alpha}_1}$. 
Our calculation shows that, for the anti-symmetric string, the ansatz (\ref{eq:ansatz_string}) gives a more severe upper bound for $T_{\bm{\mu}_2}$.

Lastly, let us consider the adjoint string $\bm{w}=\bm{\alpha}\in \Phi$. 
Applying the formula (\ref{eq:BPS_string_tension}) within the ansatz (\ref{eq:ansatz_string}), we obtain it as 
\begin{eqnarray}
T_{\bm{\alpha}}&=&{e^2 M\over \pi}\int_0^{2\pi}\diff \sigma \sqrt{(1-\cos(2\sigma))+2(N-2)(1-\cos(\sigma))}\nonumber\\
&=&{4e^2 M\over \pi}\left(\sqrt{N}+{(N-2)\over \sqrt{2}}\cosh^{-1}\sqrt{N\over N-2}\right)\nonumber\\
&=&{1\over N-1}\left(N+{(N-2)\sqrt{N}\over \sqrt{2}}\cosh^{-1}\sqrt{N\over N-2}\right) T_{\bm{\mu_1}}. 
\end{eqnarray}
According to this formula, the adjoint string tension satisfies $T_{\bm{\alpha}}\ge 2T_{\bm{\mu}_1}$, and turns out to be only slightly larger than $2T_{\bm{\mu}_1}$. 
But here it turns out that we can do a little bit better. 
Let us explicitly take $\bm{w}=\bm{\alpha}_1$, and consider a double-string ansatz, in which the adjoint string consists of two fundamental strings, $\bm{\mu}_1$ and $\bm{\alpha}_1-\bm{\mu}_1$. 
Up to permutation, $\bm{\mu}_1$ and $\bm{\alpha}_1-\bm{\mu}_1$ are related by complex conjugation, and we thus find 
\begin{equation}
    T_{\bm{\alpha}}=T_{\bm{\mu}_1}+T_{\bm{\alpha}_1- {\bm{\mu}_1}}=2T_{\bm{\mu}_1}. 
\end{equation}
Therefore, for the adjoint string, the two-independent-string ansatz is slightly better than (\ref{eq:ansatz_string}), unlike the case of anti-symmetric string. 

\section{Semi-Abelian theory}\label{sec:SemiAbelian}

In this section, we consider the $\mathrm{U}(1)^{N-1}\rtimes \mathrm{S}_N$ gauge theory obtained by gauging the $\mathrm{S}_N$ global symmetry of the $\mathrm{U}(1)^{N-1}$ gauge theory considered in Section~\ref{sec:Abelian_lattice}. We call it the semi-Abelian gauge theory.

\subsection{Gauging of the $\mathrm{S}_N$ global symmetry}\label{sec:SN_gauging}

In relativistic quantum field theories, global symmetry is generated by a set of codim-$1$ defects, which are topological and obey the group-multiplication law~\cite{Gaiotto:2014kfa}. When the global symmetry is discrete, we can gauge it by summing over all possible networks of such codim-$1$ defects. 
This procedure may be obstructed by anomalies, 
which are characterized by a topological action in one higher dimension~\cite{Cho:2014jfa, Kapustin:2014lwa, Kapustin:2014zva}. 
We also note that the gauging procedure admits the freedom to add a topological phase to each network configuration of the topological defects as long as it is consistent with locality and unitarity. 

In this section, we gauge the $\mathrm{S}_N$ symmetry of the $\mathrm{U}(1)^{N-1}$ lattice gauge theory of Section~\ref{sec:Abelian_lattice}. 
Assuming that its low-energy description enjoys emergent Lorentz symmetry due to the cubic lattice rotational invariance, the gauging procedure of $\mathrm{S}_N$ should  fit into the above general discussion. 
Absence of the $\mathrm{S}_N$ anomaly is guaranteed by the explicit construction of the lattice gauge theory. 
As extra topological terms, there are Dijkgraaf-Witten (DW) terms~\cite{Dijkgraaf:1990nc} characterized by $H^3(B \mathrm{S}_N, \mathrm{U}(1))$, which are nontrivial for all $N\ge 2$.\footnote{Although detailed information is not relevant for us as we neglect the nontrivial DW twist, let us give its full information for completeness, which may be useful for possible extensions. By the universal coefficient theorem, we obtain $H^d(B \mathrm{S}_N, \mathrm{U}(1))\simeq H^{d+1}(B \mathrm{S}_N, \mathbb{Z}) \simeq H_d (B \mathrm{S}_N, \mathbb{Z})$ because they only have the torsion part. We can find in literatures that the list of the 3d DW twist is given as 
\begin{equation}
\begin{array}{c|cccccc}
N 							  & 2 & 3 & 4 & 5 & 6 & \cdots \\ \hline
H^3(B \mathrm{S}_N, \mathrm{U}(1)) & \,\,\mathbb{Z}_2\, & \,\mathbb{Z}_6\, & \mathbb{Z}_2\oplus \mathbb{Z}_{12} & \mathbb{Z}_2\oplus \mathbb{Z}_{12} & (\mathbb{Z}_2)^2\oplus \mathbb{Z}_{12} & \cdots
\end{array}
\end{equation}
For $N\ge 6$, this group cohomology stabilizes and $H^3(B \mathrm{S}_N, \mathrm{U}(1))\simeq (\mathbb{Z}_2)^2\oplus \mathbb{Z}_{12}$, i.e. we can add three distinct DW terms, two of which give the $(\pm 1)$ phases and the another one gives the phases $\exp\left({2\pi\im\over 12}n\right)$, in the path integral of $\mathrm{S}_N$ gauge fields. }
In this paper, we limit ourselves to the case without the $3$d $\mathrm{S}_N$ DW term. 

We shall now construct the semi-Abelian gauge theory on a cubic lattice, and the simplest way to proceed is to gauge $\mathrm{S}_N$ in the Wilson formulation~(\ref{eq:Wilson_Formulation}). 
For concreteness, it is convenient to realize the semi-Abelian gauge symmetry by $N\times N$ matrices. 
We can realize an element of $\mathrm{U}(1)^{N-1}\rtimes \mathrm{S}_N$ inside $\mathrm{U}(N)$ as 
\begin{equation}
P\cdot C\in \mathrm{U}(N), 
\end{equation}
where $C=\mathrm{diag}(\rme^{\im a_1},\cdots, \rme^{\im a_N})$ with $\det(C)=1$ describes the Cartan components, and $P\in \mathrm{S}_N$ is the $N\times N$ matrix representation of a Weyl reflection, which is realized as a permutation matrix. 
The group multiplication law is given as 
\begin{equation}
(P_1\cdot C_1)(P_2\cdot C_2)= \underbrace{(P_1 P_2)}_{\in \mathrm{S}_N}\cdot \underbrace{((P_2^{-1}C_1 P_2)C_2)}_{\in \mathrm{U}(1)^{N-1}}, 
\end{equation}
and this expression elucidates the semi-direct structure in a concrete manner. 
Letting $(P_\ell\cdot C_\ell)\in \mathrm{U}(N)$ denote the link variable, the gauge-invariant plaquettes for the Lagrangian consist of two terms:
\begin{equation}
\mathrm{Re}\left(\beta_1 \tr\left[\bm{1}_N-\prod_{\ell\subset \partial p}(P_\ell\cdot C_\ell)\right]+\beta_2\tr\left[\bm{1}_N-\prod_{\ell\subset \partial p}P_\ell\right]\right). 
\label{eq:Sn_gauged_Lagrangian}
\end{equation}
The second term is the gauge-invariant kinetic term only for $\mathrm{S}_N$. By sending $\beta_2\to +\infty$, we can impose the flatness condition on the $\mathrm{S}_N$ gauge field,
\begin{equation}
\prod_{\ell \subset \partial p} P_\ell = \bm{1}_N \in \mathrm{S}_N. 
\label{eq:flatness_condition}
\end{equation}
As it is this limit that fits into the general discussion given above for the continuum description, we shall work with this flatness condition (\ref{eq:flatness_condition}) on the $\mathrm{S}_N$ link variables.  
We can readily check that the Lagrangian (\ref{eq:Sn_gauged_Lagrangian}) is equal to the Wilson action~(\ref{eq:Wilson_Formulation}) when the $\mathrm{S}_N$ gauge fields are trivial, i.e. $P_\ell=1$ for all the links $\ell$, by putting $\beta_1= \b$. 
In this manner, we obtain the $\mathrm{U}(1)^{N-1}\rtimes \mathrm{S}_N$ gauge theory out of the $\mathrm{U}(1)^{N-1}$ pure gauge theory.

Because of the flatness condition~(\ref{eq:flatness_condition}), the local dynamics should not be much affected by the gauging of $\mathrm{S}_N$, and all the interesting things in the deep infrared have to do with the global aspects of the theory. 
To be more precise, let us assume that we prepare a sufficiently large torus $T^3$ for the spacetime and that we are interested in computing correlation functions inside an open ball $B^3\subset T^3$, which has a trivial topology. 
Using the flatness condition, we may perform a local $\mathrm{S}_N$ gauge transformation so that the $\mathrm{S}_N$ gauge fields $P_\ell$ are fixed to equal 1 inside $B^3$. 
Hence, the correlation functions should be identical with those of the $\mathrm{U}(1)^{N-1}$ theory of Section~\ref{sec:Abelian_lattice} as it has a nonzero mass gap, as long as we neglect exponentially small corrections that vanish in the thermodynamic limit.  
In this sense, the gauging of $\mathrm{S}_N$ is locally trivial.

More physically, by sending the parameter $\beta_2\to \infty$ in the Lagrangian, we make the magnetic monopoles for the $\mathrm{S}_N$ gauge group extremely heavy. 
As a result, the $\mathrm{S}_N$ gauge fields are deconfined; i.e. the Wilson loops of $\mathrm{S}_N$ gauge fields obey the perimeter law at any length scale. 
Now, assume that we wish to compute the correlation functions of local operators. 
For the sake of exposition, consider a two-point function of the $\mathrm{U}(1)^{N-1}$ gauge theory
\begin{equation}
\left\langle O_1(x_1) O_2(x_2)\right\rangle_{\mathrm{U}(1)^{N-1}},
\end{equation}
though it is straightforward to extend the discussion to general $n$-point functions. As the $\mathrm{S}_N$ global symmetry is not spontaneously broken in the $\mathrm{U}(1)^{N-1}$ theory, this correlation function has a non-zero expectation value in the thermodynamic limit only if $O_1(x_1) O_2(x_2)$ contains an $\mathrm{S}_N$-singlet component. 
Thus, we may assume that $O_1(x_1) O_2(x_2)$ is $\mathrm{S}_N$ singlet without loss of generality.  
Here, we note that the operator $O_1$, $O_2$ can be $\mathrm{S}_N$ non-singlet, but they have to be mutually conjugate representations. 
By introducing an $\mathrm{S}_N$ Wilson line $W^{(\mathrm{S}_N)}(x_1,x_2)$ connecting $x_1$ and $x_2$, we can construct the $\mathrm{S}_N$ gauge invariant operator $O_1(x_1)W^{(\mathrm{S}_N)}(x_1,x_2) O_2(x_2)$. 
Since the $\mathrm{S}_N$ gauge field is deconfined, we find that $\left\langle O_1(x_1)W^{(\mathrm{S}_N)}(x_1,x_2) O_2(x_2)\right\rangle_{\mathrm{U}(1)^{N-1}\rtimes \mathrm{S}_N}$ is invariant under any continuous deformation of the path connecting $x_1$ and $x_2$. In particular, by choosing a trivial path which does not go around any nontrivial cycle of the spacetime, we obtain 
\begin{equation}
\left\langle O_1(x_1)W^{(\mathrm{S}_N)}(x_1,x_2) O_2(x_2)\right\rangle_{\mathrm{U}(1)^{N-1}\rtimes \mathrm{S}_N}=\left\langle O_1(x_1) O_2(x_2)\right\rangle_{\mathrm{U}(1)^{N-1}}
\end{equation}
in the thermodynamic limit.
Thus, any local correlation function of the $\mathrm{U}(1)^{N-1}$ theory can be recovered in the semi-Abelian gauge theory. 

\subsection{$\mathbb{Z}_N$ center symmetry}

In this and the following subsections, we discuss properties of the (electric) Wilson loops in order to identify the string tensions from the viewpoint of the symmetry. 
Here, we pay especial attention to the $1$-form symmetry, or center symmetry, of the $\mathrm{U}(1)^{N-1}\rtimes \mathrm{S}_N$ gauge theory.

The $1$-form symmetry is generated by codim-$2$ topological defects, whose fusion rule obeys group multiplication~\cite{Gaiotto:2014kfa}. 
In general, when we consider a pure gauge theory with a gauge group $G_{\mathrm{gauge}}$, the theory enjoys a $1$-form symmetry with the symmetry group $Z(G_{\mathrm{gauge}})$, which is the center of $G_{\mathrm{gauge}}$. 
Since this acts as $Z(G_{\mathrm{gauge}})$ phase rotations on Wilson loops, this has been historically called the center symmetry. 

In our case, the gauge group is $G_{\mathrm{gauge}}=\mathrm{U}(1)^{N-1}\rtimes \mathrm{S}_N$, and thus
\begin{equation}
Z(G_{\mathrm{gauge}})=\mathbb{Z}_N. 
\end{equation}
To see this, it is convenient to use the embedding of $\mathrm{U}(1)^{N-1}\rtimes \mathrm{S}_N\subset \mathrm{U}(N)$ used above, and to consider the defining representation of the latter. 
Using Schur's lemma, one sees that the $N\times N$ matrix representation of center elements must be proportional to the identity matrix. 
Such matrices are included only in $\mathrm{U}(1)^{N-1}$, which is the same as the Cartan factor of $\mathrm{SU}(N)$, and thus the center elements of $\mathrm{U}(1)^{N-1}\rtimes \mathrm{S}_N$ are the same as those of $\mathrm{SU}(N)$.  
Before gauging $\mathrm{S}_N$, the $1$-form symmetry group is given by $Z(\mathrm{U}(1)^{N-1})=\mathrm{U}(1)^{N-1}$ since the theory is an Abelian gauge theory without any electric matter fields.
Therefore, in view of the $1$-form symmetry, one might be led to claim that the semi-Abelian gauge theory should be more similar to $\mathrm{SU}(N)$ gauge theories than the $\mathrm{U}(1)^{N-1}$ theory from which it came. 

This, however, raises the following puzzle about the string tensions. In the $\mathrm{U}(1)^{N-1}$ theory, there are infinitely many different string tensions depending on the representations of Wilson loops, which are characterized by charges of the $\mathrm{U}(1)^{N-1}$ $1$-form symmetry. 
As we have seen in the previous subsection, the local dynamics is not affected by the gauging of $\mathrm{S}_N$ because we can locally set the $\mathrm{S}_N$ gauge field to be zero by gauge transformations. 
As long as we measure string tensions using large and contractible Wilson loops, the same discussion from before should apply here, and thus, there have to be infinitely many different string tensions also for the $\mathrm{U}(1)^{N-1}\rtimes \mathrm{S}_N$ gauge theory. 
But this seems rather unnatural, because the $\mathbb{Z}_N$ $1$-form symmetry is too weak to give selection rules for these string tensions. 
To put the question another way:  
\begin{quote}
\emph{How is the presence of infinitely many different string tensions compatible with the finite center symmetry?}
\end{quote} 

To make things more concrete, let us construct the generator of $\mathbb{Z}_N^{[1]}$ out of $(\mathrm{U}(1)^{[1]})^{N-1}$ generators of the $\mathrm{U}(1)^{N-1}$ gauge theory. 
In terms of the dual photon field $\bm{\sigma} \in \mathbb{R}^{N-1}/2\pi \Gamma_w$ from the monopole gas description in Section~\ref{sec:mass_gap_and_spectrum}, the generators of $(\mathrm{U}(1)^{[1]})^{N-1}$ are given by 
\begin{equation}
U_{\theta}^{(k)}(C)=\exp\left(\im {\theta\over 2\pi} \int_C \bm{\alpha}_k\cdot \diff \bm{\sigma} \right),\qquad k=1,\ldots, N-1,
\label{eq:abelian_one_form_generator}
\end{equation}
where the transformation parameter $\theta$ is $2\pi$ periodic due to the $2 \pi \G_{\mathrm w}$-periodicity of $\bm \sigma$. After gauging $\mathrm{S}_N$, these operators are no longer gauge invariant because the dual photon field $\bm{\sigma}$ transforms under the standard representation of the Weyl permutations $\mathrm{S}_N$. 
Nevertheless, the generators of $\mathbb{Z}_N^{[1]}$ can be constructed as
\begin{equation}
U_n(C) \equiv \prod_{k=1}^{N-1} U_{{2\pi\over N}kn}^{(k)}(C)= \exp\left(\im{n\over N}\int_C (\bm{\alpha}_1+2\bm{\alpha}_2+\cdots+(N-1)\bm{\alpha}_{N-1})\cdot \diff \bm{\sigma}\right). 
\end{equation}
Thanks to the periodicity of $\bm{\sigma}$, the $U_n(C)$ are invariant under $\mathrm{S}_N$ transformations and thus remain good operators for the semi-Abelian gauge theory. Moreover, we have the group multiplication law
\begin{equation} 
U_{n}(C) U_{m}(C) = U_{n+m  \; {\rm mod  \; }  N}(C)
\label{eq:one_form_generator}
\end{equation}
so $U(C) \equiv U_1(C)$ only  generates the $\mathbb{Z}_N$ subgroup of $\mathrm{U}(1)^{N-1}$. 
Other generic combinations $\prod_k U_{\theta_k}^{(k)}$ cannot satisfy the $\mathrm{S}_N$ invariance, and they drop out from the possible generators of the $1$-form symmetry. 

\subsection{String tensions beyond $N$-ality, and noninvertible topological lines}

Let us explicitly check whether or not the string tensions of the semi-Abelian gauge theory obey the standard $N$-ality rule. 
Using the embedding $\mathrm{U}(1)^{N-1}\rtimes \mathrm{S}_N\subset \mathrm{U}(N)$, we construct the Wilson loops with the $\mathrm{U}(N)$ gauge field first, and then we restrict it to the $\mathrm{U}(1)^{N-1}\rtimes \mathrm{S}_N$ gauge field.  
As we can locally eliminate the $\mathrm{S}_N$ gauge field, we may restrict the $\mathrm{U}(N)$ gauge field to its diagonal component in a naive way, as long as the Wilson loop is contractible. 
The $N$-ality of the obtained Wilson loop is the same as that of the Wilson loop with $\mathrm{SU}(N)$ gauge fields.

Let us write
\begin{equation}
W_k(C)=\exp\left(\im \Bigl(\bm{\mu}_1-\sum_{j=1}^{k-1}\bm{\alpha}_j\Bigr)\cdot \int_C \bm{A}\right), \qquad k=1,\ldots, N, 
\end{equation}
so that, for example, the fundamental Wilson loop is given by 
\begin{equation}
W_{\mathrm{fd}}(C)=W_1(C)+W_2(C)+\cdots+W_N(C). 
\end{equation}
Each Wilson loop $W_i$ in $W_{\mathrm{fd}}$ has the same string tension, and for large loops $C$ it obeys the area law: 
\begin{equation}
    \Bigl\langle W_{\mathrm{fd}}(C)
    \Bigr\rangle 
    \sim \exp(-T_{\bm{\mu}_1} \mathrm{Area}). 
\end{equation}
Under the $1$-form symmetry, 
\begin{equation}
\mathbb{Z}_N^{[1]}: W_{\mathrm{fd}}\mapsto \rme^{2\pi\im/N}W_{\mathrm{fd}},
\end{equation}
or, more precisely, 
\begin{align}
\langle U(C_1) W_{\mathrm{fd}}(C_2)\rangle = \exp\left({2\pi\im\over N}\mathrm{Link}(C_1,C_2)\right)\langle W_{\mathrm{fd}}(C_2)\rangle.
\end{align}
In order to determine whether string tensions are controlled by the $1$-form symmetry, let us consider the adjoint Wilson loop,
\begin{equation}
W_{\mathrm{ad}}(C)=\sum_{i\not =j} W_i(C) W_j^*(C)= |W_\mathrm{fd}|^2-N. 
\end{equation}
This has trivial $N$-ality, but we can readily check that its string tension is \emph{not} zero using the result of Section~\ref{sec:string_tension_Abelian}:
\begin{equation}
    \Bigl\langle W_{\mathrm{ad}}(C)
    \Bigr\rangle 
    \sim \exp(-T_{\bm{\alpha}} \mathrm{Area}),
\end{equation}
with $T_{\bm{\alpha}}\simeq 2T_{\bm{\mu}_1}$. 
This example clearly tells us that the string tensions of the semi-Abelian gauge theory carry detailed data of its gauge-group representations, which cannot be captured by the $\mathbb{Z}_N^{[1]}$ symmetry. 

Something new is needed to explain the failure of the $N$-ality rule, and this is where the non-invertible topological lines~\cite{Bhardwaj:2017xup, Buican:2017rxc, Freed:2018cec,  Chang:2018iay,Thorngren:2019iar,Ji:2019jhk, Rudelius:2020orz, Komargodski:2020mxz, Aasen:2020jwb} come in. We can easily construct such an operator by summing over all the $\mathrm{S}_N$ conjugates of $U^{(1)}_\theta(C)$: 
\begin{eqnarray}
\mathcal{T}_{\theta}(C)&\equiv&{1\over N!}\sum_{P\in \mathrm{S}_N} P U_\theta^{(1)}(C) P^{-1}\nonumber\\
&=&{1\over N(N-1)}\sum_{\bm{\alpha}\in \Phi} \exp\left(\im{\theta\over2\pi}\int_C \bm{\alpha}\cdot \diff \bm{\sigma}\right). 
\end{eqnarray}
Since this operator is $\mathrm{S}_N$ singlet, it can be a physical operator of the $\mathrm{S}_N$-gauged theory. Since each operator in the sum is topological, so too is $\mathcal{T}_\theta(C)$. Therefore, this $\mathrm{S}_N$-invariant operator shares important features with the $1$-form symmetry generators.\footnote{
Following the same logic, we can in fact construct many more continuous families of non-invertible symmetries. Indeed, we can average over all $\mathrm{S}_N$ conjugates of an arbitrary product $U_{\theta_1}^{(k_1)}(C) \cdots U_{\theta_r}^{(k_r)}(C)$ of the operators \eqref{eq:abelian_one_form_generator} to get a non-invertible symmetry generator $\mathcal{T}_{\theta_1, \ldots , \theta_r}^{(k_1, \ldots, k_r)}(C)$. In particular, the non-invertible symmetry generator $\mathcal{T}_{\theta_1 , \ldots , \theta_{N-1}}^{(1, \ldots, N-1)}(C)$ with $\theta_k = 2 \pi k /N$ will coincide with the $\mathbb{Z}_N^{[1]}$ center symmetry generator $U(C)$, as one can readily check from \eqref{eq:one_form_generator}. Thus, the $\mathbb{Z}_N^{[1]}$ center symmetry is actually contained within a continuous family of non-invertible symmetries.
} 
However, the group multiplication law is not satisfied for $\mathcal{T}_\theta(C)$, as one can easily check: 
\begin{equation}
    \mathcal{T}_{\theta}(C)\mathcal{T}_{\theta'}(C)\not= \mathcal{T}_{\theta+\theta'}(C). 
\end{equation}
Because of the violation of the group multiplication property, we cannot regard $\mathcal{T}_{\theta}(C)$ as a generator of an ordinary $1$-form symmetry in contrast with (\ref{eq:one_form_generator}).
Instead, it is a generator of a non-invertible symmetry.

Let us consider the component of the Wilson loop that corresponds to the weight $\bm{w}\in \Gamma_\mathrm{w}$. Its eigenvalue for $\mathcal{T}_\theta$ is given by 
\begin{equation}
    {1\over N(N-1)}\sum_{\bm{\alpha}\in\Phi}\exp\left(\im \theta \bm{\alpha}\cdot \bm{w}\right). 
\end{equation}
As a consequence, the fundamental Wilson loop transforms as 
\begin{equation}
    W_{\mathrm{fd}}\mapsto {1\over N}\bigl(N-2+2\cos(\theta)\bigr) W_{\mathrm{fd}}.
\end{equation}
More importantly, the adjoint Wilson loop also transforms nontrivially as 
\begin{equation}
    W_{\mathrm{ad}}\mapsto {(N-2)(N-3)+4(N-2)\cos(\theta)+2\cos(2\theta)\over N(N-1)}W_{\mathrm{ad}}. 
\end{equation}
This elucidates that we can detect the detailed information of the Wilson loop beyond $N$-ality by using the non-invertible topological line operator $\mathcal{T}_\theta$. 

As another example, we can detect the difference between the symmetric and anti-symmetric two-index representations, $W_{\mathrm{sym}}$ and $W_{\mathrm{asym}}$, of $\mathrm{SU}(N)$, whose highest weights are given by $2\bm{\mu}_1$ and $\bm{\mu}_2$, respectively. 
We have to note, however, that $W_{\mathrm{sym}}$ is not an eigen-operator of $\mathcal{T}_{\theta}$, because the two-index symmetric representation of $\mathrm{SU}(N)$ decomposes into two irreducible representations of $\mathrm{U}(1)^{N-1}\rtimes \mathrm{S}_N$. 
Since $\bm{\mu}_2=2\bm{\mu}_1-\bm{\alpha}_1$ and $(2\bm{\mu}_1)\cdot  \bm{\alpha}_1=2$, each charge in the anti-symmetric representation appears in the list of charges of the symmetric representation exactly once,  
and thus the correct eigen-operator is $W_{\mathrm{sym}}-W_{\mathrm{asym}}$. 
Indeed, one can check that 
\begin{equation}
    \Bigl\langle (W_{\mathrm{sym}}-W_{\mathrm{asym}})(C)
    \Bigr\rangle 
    \sim \exp(-2T_{\bm{\mu}_1} \mathrm{Area}),\qquad 
    \Bigl\langle W_{\mathrm{asym}}(C)) \Bigr\rangle 
    \sim \exp(-T_{\bm{\mu}_2} \mathrm{Area}), 
\end{equation}
with $T_{\bm{\mu}_2}<2T_{\bm{\mu}_1}$, as we have discussed in Section~\ref{sec:string_tension_Abelian}. 
We find 
\begin{equation}
(W_{\mathrm{sym}}-W_{\mathrm{asym}})\mapsto {N-2+2\cos(2\theta)\over N} (W_{\mathrm{sym}}-W_{\mathrm{asym}}), 
\end{equation}
and 
\begin{equation}
    W_{\mathrm{asym}}\mapsto 
    {(N-2)(N-3)+2+4(N-2)\cos(\theta)\over N(N-1)}W_{\mathrm{asym}}. 
\end{equation}
This gives another explicit demonstration of the fact that one can distinguish different string tensions for representations of the same $N$-ality with the help of the topological operator $\mathcal{T}_\theta$.

\subsection{Effect of dynamical electric particles}

In the previous section, we discussed the behavior of string tensions for the pure semi-Abelian gauge theory. 
String tensions do not obey the $N$-ality rule, and the presence of non-invertible topological lines explain why they carry more detailed information. 
In this section, we discuss what will happen to the string tensions once dynamical electric charges are added. 

Once electric charges are incorporated as dynamical excitations, their pair creation can break confining strings if it is energetically favorable. 
If the fundamental electric charge is added, we expect that all the confining strings can be broken and all Wilson loops will obey the perimeter law. 
If the adjoint charge is added instead, we expect that the string tensions should obey the $N$-ality rule, because the adjoint Wilson loop would then obey the perimeter law. 
Can we justify these expectations from the viewpoint of topological lines? 

For this purpose, we need to identify which line operators cease to be topological once the dynamical electric charges are included. 
If a line acts nontrivially on the Wilson loop corresponding to the dynamical excitations, then it is no longer topological after introducing dynamical charges~\cite{Rudelius:2020orz}.
This is because the corresponding Wilson loop can end on charged local operators, so that the linking number is no longer well-defined; in other words, the topological invariance of the symmetry operator is lost. 

Let us add dynamical adjoint particles, and then determine whether or not $\mathcal{T}_\theta$ is topological. 
Since the eigenvalue of $W_{\mathrm{adj}}$ has to be $1$ for any topological operator, $\mathcal{T}_{\theta}$ is topological only if 
\begin{equation}
    {(N-2)(N-3)+4(N-2)\cos(\theta)+2\cos(2\theta)\over N(N-1)}=1. 
\end{equation}
This is solved only by $\theta=0$ mod $2\pi$, and thus only the trivial one $\mathcal{T}_{\theta=0}=1$ is topological.
This implies that the non-invertible symmetry ceases to be an exact symmetry once an adjoint matter field is added. 
On the other hand, the $\mathbb{Z}_N$ $1$-form symmetry is kept intact because the generator acts trivially on $W_{\mathrm{ad}}$. 
In this case, the string tensions obey the $N$-ality rule at least if the Wilson loops are sufficiently large, which is consistent with the observation for $3$d $\mathrm{SU}(N)$ Yang--Mills theory. 

As a nontrivial exercise, we can add dynamical particles corresponding to $n\Gamma_{\mathrm{r}}$ with $n>1$, instead of $W_{\mathrm{ad}}$. 
Then the non-invertible line $\mathcal{T}_\theta$ is topological if 
\begin{equation}
    {(N-2)(N-3)+4(N-2)\cos(n\theta)+2\cos(2n\theta)\over N(N-1)}=1,
\end{equation}
and this has nontrivial solutions, $\theta\in (2\pi/n)\mathbb{Z}$. 
Therefore, the continuous part of the non-invertible symmetry $\mathcal{T}_{\theta}$ is explicitly broken by dynamical electric charges $n\Gamma_{\mathrm{r}}$, but the discrete part $\mathcal{T}_{\theta=2\pi k/n}, \; k=1, \ldots, n $ still generates a good non-invertible symmetry. 
As a result, Wilson lines distinguished by $\mathcal{T}_{\theta=2\pi k/n}$ can have different string tensions even if they share the same $N$-ality.

\section{Summary and Discussions}

In this paper, we have studied the properties of the semi-Abelian gauge theory in $3$ spacetime dimensions, where the gauge group is $G_{\mathrm{gauge}}=\mathrm{U}(1)^{N-1}\rtimes \mathrm{S}_N$. 
As we have imposed the flatness condition on the $\mathrm{S}_N$ gauge field, we can locally eliminate it completely, so the spectral properties of the mass gap and string tensions can be calculated as the $\mathrm{U}(1)^{N-1}$ theory. 
We have seen that the mass gap is generated via the Polyakov mechanism as a consequence of monopole-instanton proliferation. 
We can classify their magnetic charges using the $\mathrm{SU}(N)$ representation, and all the monopoles for the roots give equally dominant contributions to the effective potential. 
This point is very different from the Polyakov model or QCD(adj) with an $S^1$ compactification, where only the monopoles associated with the simple roots play the dominant role, and it comes from the $\mathrm{S}_N$ invariance of our model. 

Using the dual formulation, we also computed various string tensions, and we found that there are infinitely many different string tensions. 
When the $\mathrm{S}_N$ symmetry is not gauged, this can be explained very naturally in the context of the $1$-form symmetry because the center of $\mathrm{U}(1)^{N-1}$ is $\mathrm{U}(1)^{N-1}$ itself, and thus the $1$-form symmetry group is large enough to explain the selection rules between infinitely many confining strings. 

A puzzle arises, however, after gauging $\mathrm{S}_N$, because the center symmetry is just $Z(G_{\mathrm{gauge}})=\mathbb{Z}_N$. 
This is because most of the elements of $\mathrm{U}(1)^{N-1}$ do not commute with the permutations, and the permutation invariance requires that the center elements be proportional to the identity matrix. 
Thus, the $1$-form symmetry of semi-Abelian gauge theory is as small as that of $\mathrm{SU}(N)$ Yang--Mills theory, where the string tensions are characterized by $N$-ality alone. 
Therefore, for the semi-Abelian theory, there is a clear discrepancy between the actual behavior of the string tensions and the natural expectation from $\mathbb{Z}_N$ center symmetry. 

We find that the discrepancy is resolved by recognizing the presence of noninvertible symmetry. 
We constructed the topological line operator $\mathcal{T}_\theta$ out of the $\mathrm{U}(1)^{N-1}$ $1$-form symmetry generators, which remain well-defined and topological after gauging $\mathrm{S}_N$ but do not satisfy the group multiplication law. 
Though this operator is noninvertible, its action on the Wilson lines yield eigenvalues that are able to distinguish representations with the same $N$-ality. 
Thus, we have demonstrated the utility of an extended notion of symmetry in a $3$d toy example of a gauge theory. 

We should mention that the formal development of non-invertible symmetry is still an important task. 
In the case of higher-form or higher-group symmetry, their formalization not only provided the rigorous definition and generalization of the center symmetry, 
but also gave new tools to analyze interacting QFTs, such as generalizations of anomaly matching~\cite{Gaiotto:2017yup, Tanizaki:2017bam, Komargodski:2017smk, Shimizu:2017asf, Gaiotto:2017tne, Tanizaki:2017qhf, Tanizaki:2017mtm, Tanizaki:2018wtg, Komargodski:2017dmc, Kikuchi:2017pcp, Tanizaki:2018xto, Karasik:2019bxn, Cordova:2019jnf}. 
It would be very nice if this repertoire of useful tools could be enhanced to include non-invertible symmetry. 

Lastly, let us present some  speculation. 
As we stated in the introduction, a similar behavior regarding the $N$-ality rule has been observed in simulations of $\mathrm{SU}(N)$ Yang--Mills on the lattice: there is an intermediate distance scale where the quark-antiquark potential exhibits linear confinement but its string tension depends on the details of the gauge-group representation. 
Though it is widely believed that the string tension becomes solely dictated by $N$-ality once the quark-antiquark separation becomes sufficiently large, it is logically possible that `sufficiently large' is parametrically larger than the strong length scale $\Lambda^{-1}$ at which confinement sets in.  For instance, viewing $N$ as a parameter, it may very well be that the $N$-ality rule sets in at a distance scale $h(N) \Lambda^{-1} \gg \Lambda^{-1}$, where $h(N) \rightarrow \infty$ as $N \rightarrow \infty$. 
We think it would be an intriguing possibility if, even in pure Yang--Mills, some approximate notion of non-invertible symmetry could be used to explain the behavior of string tensions beyond $N$-ality at these intermediate distances.  

A more striking example may be QCD with fundamental or two-index matter fields, 
where the 1-form $\Z_N$ center symmetry is either completely or partially lost, or Yang--Mills theories with simply-connected gauge groups without a center, such as $G_2$.
Even in cases where 1-form symmetry is completely lost, we believe that an approximate  non-invertible symmetry could potentially give a precise meaning  to confinement of arbitrary test charges, and therefore provide the long sought definition of confinement in such theories.

\acknowledgments

We thank Shailesh Chandrasekharan, Hanqing Liu, Hersh Singh, Misha Shifman, Mike Creutz, Tin Sulejmanpasic, and Aleksey Cherman   for useful discussions. 
The authors thank the YITP–RIKEN iTHEMS workshop ``Potential Toolkit to Attack Nonperturbative Aspects of QFT –Resurgence and related topics–'' (YITP-T-20-03) for providing opportunities of useful discussions.
The work of Y.~T. was partially supported by  JSPS KAKENHI Grant-in-Aid for Research Activity Start-up, 20K22350.  
M.~\"U.  is supported by the U.S. Department of Energy, Office of Science, Division of Nuclear Physics under Award DE-SC0013036.  

\appendix 

\section{Review of Abelian duality on the lattice}
\subsection{Differential forms on the lattice}

There is, on the lattice, a close analog of the notion of differential forms, and it is especially convenient for treating Abelian lattice gauge theories. Here we give a somewhat informal introduction to this formalism, which we use throughout this article. See \cite{Sulejmanpasic:2019ytl} for a more systematic discussion. 

We begin with a bit of lattice geometry. Consider a $d$-dimensional cubic lattice $\Lambda^d$. Such a lattice contains `$r$-cells' $c^{(r)}$ for each $r=0,1,\ldots,d$. 
Thus, the $0$-cells are the sites $s$, the $1$-cells the links $\ell$, the $2$-cells the plaquettes $p$, the $3$-cells the cubes $c$, etc., and everything is oriented. 
For example, for a link $\ell = (x ; \hat{\m})$, the oppositely oriented link is given by $-\ell \equiv (x + \hat{\m}; - \hat{\m})$, and these are to be viewed as distinct objects despite corresponding to the same unoriented edge.\footnote{In the case of sites, an opportunity for confusion may arise. In this notation, the sites $s$ and $-s$ correspond to the same point $x$, say, but are equipped with opposite orientations.}
By convention, whenever we write a sum or product over $r$-cells, we do not double count $r$-cells that only differ by orientation. 

The `boundary operator' $\del$ takes an $r$-cell into the (oriented) sum of the $(r-1)$-cells that constitute its boundary. For example, the boundary operator on a plaquette $p = (x ; \hat{\m}, \hat{\n})$ yields
\begin{equation}
        \del (x;  \hat{\m},  \hat{\n}) 
        = (x; \hat{\m}) + (x + \hat{\m}; \hat{\n}) - (x + \hat{\n} ; \hat{\m}) - (x; \hat{\n}).
\end{equation}
Importantly, the boundary operator is nilpotent, $\del^2 = 0$. 
By a slight abuse of notation, we write
\begin{equation}
    c^{(r-1)} \subset \del c^{(r)}
\end{equation}
if the $r$-cell $c^{(r)}$ contains in its boundary the $(r-1)$-cell $c^{(r-1)}$. 
We thus have (tautologically)
\begin{equation}
    \del c^{(r)} = \sum_{c^{(r-1)} \subset \del c^{(r)}} c^{(r-1)}.
\end{equation}

\begin{figure}
\vspace{-2cm}
\begin{center}
\includegraphics[width = 0.5\textwidth]{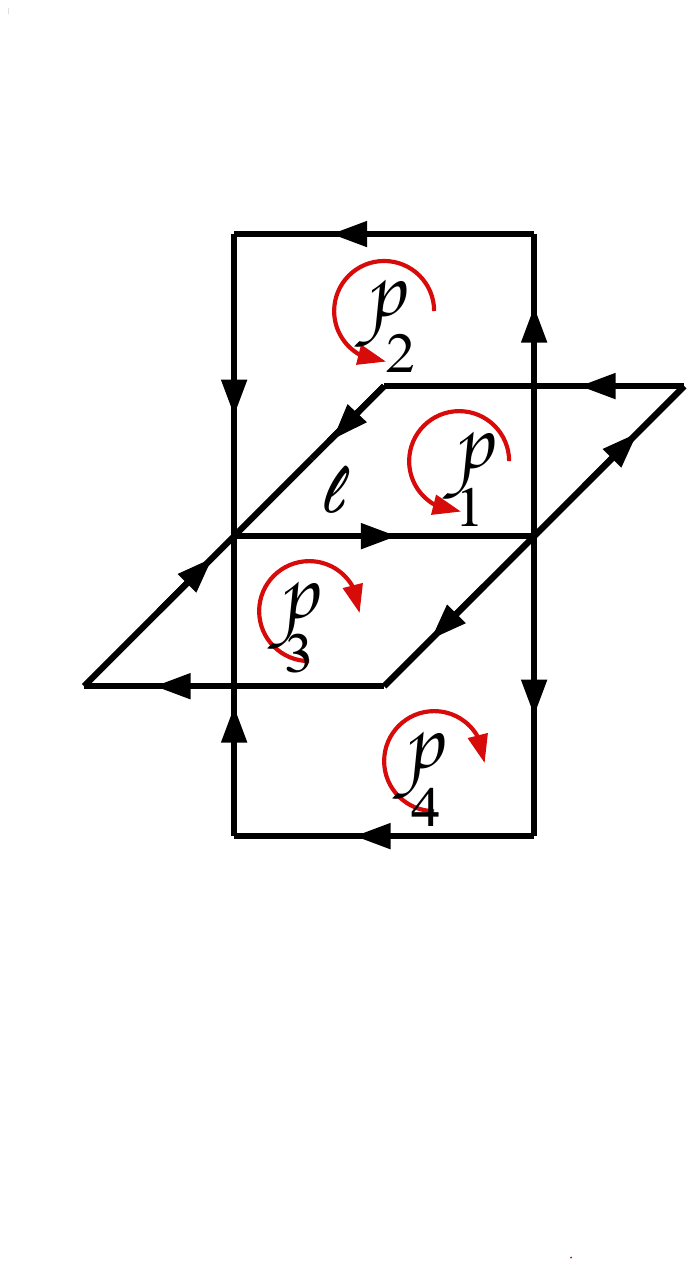}
\vspace{-4.0cm}
\caption{Coboundary operator on a link in a 3d lattice: $\delta \ell = p_1 + p_2 + p_3 + p_4$}
\label{fig:coboundary0}
\end{center} 
\vspace{-0.5cm}
\end{figure}

We also have a kind of dual to the boundary operator, the `coboundary operator' $\delta$. It takes an $r$-cell into the sum of the $(r+1)$-cells that each contains $c^{(r)}$ in its boundary,
\begin{equation}
    \delta c^{(r)} = \sum_{\del c^{(r+1)} \supset c^{(r)}} c^{(r+1)}.
\end{equation}
For example, for a link $\l = (x; \hat{1})$ in a three-dimensional lattice, we have 
\begin{equation}
    \delta (x; \hat{1}) = (x; \hat{1},\hat{3}) + (x; \hat{1},\hat{2}) + (x-\hat{3};\hat{3},\hat{1}) + (x-\hat{2};\hat{2},\hat{1})
\end{equation}
(see Figure~\ref{fig:coboundary0}).
It is easy to show that $\delta^2 = 0$.

There is another lattice $\Tilde{\L}^d$, the `dual lattice', that is naturally associated with the primary lattice $\L^d$. 
The points of $\Tilde{\L}^d$ are given by $\xt = x + \frac{1}{2}(\hat{1} + \cdots + \hat{d} )$, with $x$ any point of $\L^d$. 
These lattices are connected by an operator $*$, which takes $r$-cells in the primary lattice into $(d-r)$-cells in the dual lattice and vice versa; it is defined as follows: for an $r$-cell  $c^{(r)}$ in the primary lattice, $* c^{(r)}$ is the unique $(d-r)$-cell in the dual lattice such that $c^{(r)}$ and $*c^{(r)}$ intersect transversally, and such that the orientation of the ordered pair $(c^{(r)}, *c^{(r)})$ is positive. For example, for a plaquette $p = (x; \hat{1} , \hat{2})$ in a 3d lattice $\L^3$, we have $*p = (\xt - \hat{3} ; \hat{3})$ (see Figure~\ref{fig:poincare}). 
On $r$-cells, we have
\begin{equation}
    *^2 = (-)^{r(d-r)}.
\end{equation}

We can now define differential forms on the lattice. An $r$-form $\o$ is simply a gadget that assigns a value $\o_{c^{(r)}}$ to each $r$-cell $c^{(r)}$, and it extends as a linear map. To compare with more conventional lattice field theory notation, consider for example a 1-form $\th$. We may write its value on a link $\ell = (x; \hat{\m})$ as 
\begin{equation}
    \th_\ell \equiv \th_\m (x).
\end{equation}
We define the `exterior differential' operator $\dd$ to take $r$-forms to $(r+1)$-forms according to the formula
\begin{equation}
    (\dd \o)_{c^{(r+1)}} \equiv \sum_{c^{(r)} \subset\del c^{(r+1)}} \o_{c^{(r)}} = \o_{\del c^{(r+1)}}. 
\end{equation}
To again compare with more conventional notation, we note that the differential $\dd \th$ of the 1-form $\th$ on a plaquette $p = (x; \hat{\m} ,\hat{\n})$ is given by
\begin{equation}
    (\dd \th)_p = \theta_{\mu}({x}) +   \theta_{\nu}({x + \hat{\m} }) -  \theta_{\mu}({x+ \hat{\n}} ) - \theta_{\nu}({x}).
\end{equation}

We also define the dual $\dd^{\dagger}$ of the exterior differential, the `codifferential', which takes $r$-forms to $(r-1)$-forms, according to the formula
\begin{equation}
    (\dd^{\dagger} \o)_{c^{(r-1)}} \equiv \sum_{c^{(r)} \subset\delta c^{(r-1)}} \o_{c^{(r)}} = \o_{\delta c^{(r-1)}}.
\end{equation}
It is easy to see that $\dd^2 = (\dd^{\dagger})^2 = 0$. The star operator $*$ takes $r$-forms on the primary lattice to $(d-r)$-forms on the dual lattice, and vice versa, according to the formulae
\begin{equation}
    (* \o)_{\tilde{c}^{(d-r)}} = \o_{*\tilde{c}^{(d-r)}} \,, \qquad (*\tilde{\o})_{c^{(d-r)}} = \tilde{\o}_{*c^{(d-r)}}.
\end{equation}
It is easy to show that on $r$-forms, we have
\begin{equation}
    *^2 = (-)^{r(d-r)}.
\end{equation}

One of the more useful features of lattice form notation is that it enables us to `integrate by parts' mindlessly. That is, we have the formula
\begin{equation}
    \sum_{c^{(r)}} (\dd \o)_{c^{(r)}} \t_{c^{(r)}} = \sum_{c^{(r-1)}} \o_{c^{(r-1)}} (\dd^{\dagger} \t)_{c^{(r-1)}} .
    \label{eq:partial_integration}
\end{equation}
Actually, it is this partial integration formula that justifies calling $\dd^{\dagger}$ the dual of $\dd$.
To illustrate the utility of the notation, let us prove  (\ref{eq:partial_integration}):
\begin{equation}
        \mathrm{L.H.S.}
        \equiv \sum_{c^{(r)}} \sum_{c^{(r-1)} \subset\del c^{(r)}} \o_{c^{(r-1)}} \t_{c^{(r)}} 
        = \sum_{c^{(r-1)}} \sum_{c^{(r)} \subset\delta c^{(r-1)}} \o_{c^{(r-1)}} \t_{c^{(r)}} 
        \equiv \mathrm{R.H.S}.
\end{equation}

Finally, let us discuss the lattice analog of the `Hodge decomposition'. As in the continuum, we define the Laplacian on forms by $\D = \dd \dd^{\dagger} + \dd^{\dagger} \dd$. In particular, on 0-forms $\f$, we have 
\begin{equation}
    (\Delta \f)(x) = \sum_{\m = 1}^d \large[ 2 \f(x) - \f(x + \hat{\m}) - \f(x - \hat{\m})  \large].
\end{equation}
Forms annihilated by $\D$ are called `harmonic'. It is simple to show that harmonic forms are annihilated by both $\dd$ and $\dd^{\dagger}$. The Hodge decomposition is the statement that any $r$-form $\o$ can be written uniquely as
\begin{equation}
    \o = \dd \a + \dd^{\dagger} \b + \h,
\end{equation}
where $\h$ is harmonic. We will not prove this here.

\subsection{3d compact QED on the lattice}
\label{SecAbelianDuality}

Here we discuss the dual representation of 3d $\mathrm{U}(1)$ lattice gauge theory following the presentation of  Ref.~\cite{Gopfert:1981er} (see also \cite{Banks:1977cc,Savit:1977fw}). We also give some attention to global issues involving the spacetime topology. 
We note that, although we restrict our presentation to three dimensions, many techniques used here are also applicable in four-dimensional spacetime lattices, where interesting phase diagrams have been expected through electromagnetic dualities~\cite{Creutz:1979kf, Creutz:1979zg, Cardy:1981qy, Cardy:1981fd, Honda:2020txe}. 

We start from the Wilson formulation of the $\mathrm{U}(1)$ lattice gauge theory in $d=3$ spacetime dimensions: 
\begin{align}
\exp(-S)&=  \exp \left(     \beta   \sum_p  (\cos f_p - 1) \right)
= \prod_p
\ee^{   \beta (\cos f_p -1)}. 
\label{eq:SW1}
\end{align}
Since the action is periodic in  $f_p$, we can expand  $\exp(-S)$  as a Fourier series:
\begin{align}
\ee^{  \beta (\cos f_p -1)}  =  \sum_{k_p \in \mathbb Z}  \ee^{ \ii k_p f_p  } I_{k_p } (\beta) \ee^{-\b},
\label{Fourier} 
\end{align} 
where $I_{k_p} (\beta) $ is the modified Bessel function of the first kind of order $k_p$.   This  representation is    useful because  it  allows  us to integrate over the link fields in the Abelianized theory in a straightforward manner.

The partition function can be rewritten as 
\begin{equation}
Z= \int_0^{2 \pi} [\dd a_\l]   \exp(-S) 
 = \sum_{\{k_p \in \mathbb Z\}} \int_0^{2 \pi} [\dd a_\l]   
 \exp \left( \ii \sum_p k_p f_p \right) \prod_p I_{k_p} (\beta) \ee^{-\b}.
\label{pt4}
\end{equation}
From this expression, in the weak coupling limit $\beta\gg 1$, we can obtain the Villain form. 
Using the asymptotic expansion, $ \ee^{-\beta} I_{k_p} (\beta)   \sim \frac{1}{\sqrt{2 \pi \beta} } \ee^{- \frac{1}{2 \beta}k_p^2}$, we can rewrite the summation over $k_p$ by Poisson summation formula,
\begin{align}
\sum_{k_p \in \mathbb Z}  \ee^{ \ii k_p f_p }   \frac{1}{\sqrt{2 \pi \beta} } \ee^{- \frac{1}{2 \beta}k_p^2} 
=   \sum_{n_p \in \mathbb Z} \ee^{- \frac{\beta}{2} (f_p -2 \pi n_p )^2}
\label{pt5}
\end{align}
Here, $n_p $ can be viewed as the flux passing through the corresponding surface $p$. 
The total flux passing through the surface of the cube $c$ centered at $\xt$ is 
\begin{align}
 \oint_{\partial c} n =  q(\xt), 
 \end{align} 
 which is just the magnetic charge located at  $\xt$.
 In the following, we concentrate only on this weak-coupling limit that is exactly equivalent to the Villain formulation.

\paragraph{Dual formulation, from $\Lambda^3$ to $\Tilde{\Lambda}^3$:}
In order to obtain the dual representation of the Villain form, we perform the exact integration over $a_\ell$ before the summation over $k_p$ in (\ref{pt4}). 
As $\sum_p k_p (\diff a)_p=\sum_\ell (\diff^\dagger k)_\ell a_\ell$, the exact integration over $a_\l$ enforces the constraint,
 \begin{align}
 (\dd^{\dagger} k)_\l = 0.
 \label{constraint}
 \end{align}
As a result, the partition function can be written as a constrained sum over the $k_p$:
 \begin{align}
Z =  \sum_{\{k_p \in \mathbb Z\}} \left\{ \prod_{\l} \delta_{(\dd^{\dagger} k)_\l , 0} \right\}  \left\{ \prod_p \ee^{- \frac{1}{2 \beta}k_p^2} \right\}.
\end{align} 
To construct the dual formulation of the theory, it is useful to turn the constrained sum into an unconstrained sum.  
To this end, we consider the decomposition of  $(*k)_{\tilde{\ell}}$ as 
\begin{equation}
    (*k)_{\tilde{\ell}}=(\diff m)_{\tilde{\ell}} + \tilde{a}_{\tilde{\ell}},
     \label{duality}
\end{equation}
where $m_{\st}$ is an integer-valued scalar field,  and $\tilde{a}_{\tilde{\ell}}$ is an integer-valued link field on the dual lattice. 
Since $k_p$ satisfies the constraint \eqref{constraint}, $\tilde{a}$ can be regarded as a flat connection. 
In computing the partition function, we may make the replacement, 
$
 k_p  \rightarrow (* \dd m)_p + (* \tilde{a})_p,
$
and the constrained sum over $\{ k_p \in \mathbb Z \}$ becomes an unconstrained sum over $\{ m_{\st} \in \mathbb Z\}$ and $[\tilde{a}] \in H^1(\Tilde{\Lambda}, \mathbb{Z})$. 
As a result, the partition function in the weak-coupling limit takes the simple form, 
\begin{equation}
    Z = 
    \sum_{[\tilde{a}]\in H^1}
 \sum_{\{m_{\st} \in \mathbb Z\} }
   \exp \left( - \frac{1}{2 \beta}  \sum_{\lt}  ( \dd m + \tilde{a})_{\lt}^2   \right),
   \label{dual-def}
\end{equation}
after using the duality relation (\ref{duality}). 
 This model is sometimes called the
 ${\mathbb Z}$-ferromagnet when $\tilde{a}=0$.

Two more steps are needed to convert $\Z$-ferromagnet representation to a continuum QFT. 
 First,  convert the sum into an integration over a continuous variable.  Using  the Poisson resummation identity repeatedly through the lattice    $ \Tilde{\Lambda}_3 $, 
 \begin{align}
 \sum_{m(\xt ) \in \mathbb Z} \delta(\sigma(\xt )- 2\pi m(\xt )) =    \sum_{q(\xt ) \in \mathbb Z} \ee^{ \ii\, q(\xt ) \sigma(\xt )}, 
  \end{align}
we immediately obtain the partition function as an  infinite dimensional integral,
\begin{align}
Z& = 
\sum_{[\tilde{a}]\in H^1}
 \int  [ \dd \sigma (\xt)]   \;  \sum_{ \{q(\xt)\} }  \; 
  \exp \left(- \frac{1}{8\pi^2 \beta}  \sum_{\xt } ( \partial_{\mu}^{-} \sigma({\xt})+2\pi \tilde{a}_\mu(\tilde{x}))^2
 + \ii \sum_{\xt } q({\xt }) \sigma({\xt })
    \right) ,
    \end{align}
where $q({\xt }) \in \Z$ has an interpretation as the magnetic charge of a monopole-instanton at position ${\xt } \in   \Tilde{\Lambda}_3 $.
The kinetic term of this expression clarifies that $\tilde{a}$ plays the role of the gauge field for the discrete shift symmetry $\sigma\mapsto \sigma+2\pi$, and thus the dual photon field $\sigma$ is $2\pi$-periodic scalar. 
Having made this point, for simplicity of notation, we shall neglect the effect of nontrivial topology from now on, and set $\tilde{a}=0$. 

The Gaussian integration over $\sigma$ can be done exactly to produce the  Coulomb gas representation for the magnetic monopoles:
\begin{align}
Z& = 
 \sum_{ \{q(\xt)\} } 
\exp{ \left( \frac{1}{2 \beta}  \sum_{\xt , \xt'} (- 4 \pi^2 \beta^2 ) q({\xt }) v(  \xt  - \xt ')  q({\xt '}) \right)},    
    \end{align}
where $v(  \xt  - \xt ')$  is the three dimensional    Coulomb  interaction formulated on the lattice (lattice Green function), formally given by $v(\xt ) = \Delta^{-1}$. 
Let us split this Green function into two parts by adding and subtracting $ (\Delta +M_{\mathrm{PV}}^2)^{-1} $,
\begin{align}
\Delta^{-1} &=   \Delta^{-1} - (\Delta +M_{\mathrm{PV}}^2)^{-1}+  (\Delta +M_{\mathrm{PV}}^2)^{-1}   \cr
& =  \Delta^{-1} ( 1 + \Delta/M_{\mathrm{PV}}^2)^{-1}  +  (\Delta +M_{\mathrm{PV}}^2)^{-1}    \cr
& =  u_{M_\mathrm{PV}}(\xt )  + w_{M_{\mathrm{PV}}}( \xt ) ,
\end{align}
where $u_{M_{\mathrm{PV}}}(\xt )$ is the Green function of the Pauli-Villars (PV) regulated Laplacian $\Delta_{M_{\mathrm{PV}}} \equiv \Delta (1 + \Delta/M_{\mathrm{PV}}^2)$, and  $w_{M_{\mathrm{PV}}}( \xt )$ is the Yukawa  Green function.

With this decomposition, we reintroduce the scalar field $\sigma$ for the PV regulated propagator $u_{M_{\mathrm{PV}}}(\tilde{x})$, and then we obtain the partition function as
\begin{equation}
Z = 
 \int  [ \dd \sigma (\xt)]   \;  \sum_{ \{q(\xt)\} }  \; 
  \ee^{ \sum_{\xt }\left(- \frac{1}{8\pi^2 \beta}    
   \sigma({\xt} )\Delta_{M_{\mathrm{PV}}}   \sigma({\xt } ) +  \ii\, q({\xt }) \sigma({\xt })\right) 
   - 2\pi^2\beta \sum_{\xt , \xt '} q({\xt }) w_{M_{\mathrm{PV}}}(  \xt  - \xt ')  q({\xt '})  }.
\end{equation}
Since $w_{M_{\mathrm{PV}}}(  \xt  - \xt ')$ is a massive propagator with the PV mass $M_{\mathrm{PV}}$, it is exponentially damping if $|\tilde{x}-\tilde{x}'|\gtrsim 1/M_{\mathrm{PV}}$. 
Therefore, we can take $w_{M_{\mathrm{PV}}}(  \xt  - \xt ')  = w_{M_\mathrm{PV}}(0) \delta_{\xt  - \xt ', 0}$, where $w_{M_{\mathrm{PV}}}(0) = v(0) - {\cal O}(1/M_{\mathrm{PV}}) \approx  0.253 - {\cal O}(1/M_{\mathrm{PV}})  $. 
In this limit, the partition function simplifies into
\begin{equation}
Z = 
 \int [ \dd \sigma(\xt)]  
  \ee^{ - \frac{1}{8\pi^2\beta}\sum_{\xt} 
   \sigma({\xt }) \Delta_{M_{\mathrm{PV}}}    \sigma({\xt } ) }
   \sum_{\{q(\xt)\} } 
   \prod_{\xt}\rme^{- 2 \pi^2 \beta  v(0) (q({\xt }))^2+\ii q({\xt }) \sigma({\xt }) } , 
\label{pf}
    \end{equation}
In \eqref{pf}, $ 2 \pi^2 \beta  v(0) (q({\xt }))^2  $  has an interpretation as the action of the configurations  with magnetic charge $q({\xt })$. Let us denote the minimal action by $S_0=2\pi^2 \beta v(0)$.  

We now perform the dilute-gas approximation as the leading-order semiclassical approximation. 
We only take into account the minimal effect of the monopole-instantons corresponding to $q(\xt)=\pm 1$ seriously, and regard higher-order effects in $\rme^{-S_0}$ as unimportant. As a result, we may approximate the sum over $q(\xt)$ by
\begin{equation}
    \sum_{q(\xt)} \rme^{- S_0 q(\xt)^2+\ii q(\xt) \sigma({\xt })}= \exp\left(2\rme^{-S_0} \cos(\sigma({\xt }) )\right) + \mathcal{O}(\rme^{-2S_0}). 
\end{equation}
Substituting this expression into (\ref{pf}), we obtain the local Lagrangian for the dual photon field:
\begin{equation}
    Z = \
 \int [ \dd \s (\xt)]  
  \exp\left( - \frac{1}{8\pi^2\beta} \sum_{\xt} \s({\xt }) \Delta_{M_{\mathrm{PV}}}   \s({\xt } )
  + 2\rme^{-S_0} \sum_{\xt} \cos(\s(\xt))\right).
\end{equation}

\section{Wilson to Villain at weak coupling} \label{SecWilsonFormulation}

As mentioned in Section~\ref{SecDescription}, semi-Abelian ${\rm U}(1)^{N-1}$ gauge theory may also be given in the Wilson formulation by taking the action  
\begin{equation}
    S_{\rm W} = \b \sum_p \sum_{i=1}^N (1 - \cos f^i_p) - \ii \sum_\l \sum_{i=1}^N v_\l a^i_\l,
\end{equation}
where the $a^i_\l \in [0,2 \pi]$ are ${\rm U}(1)$ gauge fields, the $f^i_p = (\dd a^i)_p$ are the corresponding field strengths, and $v_\l \in \Z$ is a Lagrange multiplier.
This expression has manifest ${\rm S}_N$ global symmetry. 
The purpose of this appendix is to demonstrate the weak-coupling equivalence of this formulation and the Villain one \eqref{multiVillain}.

The first step is to use  \eqref{Fourier} for $\cos(f_p^i)$ for $i=1,\ldots, N$ with the weak-coupling approximation, and then to apply \eqref{pt5} for $i=1,\ldots, N-1$.  We obtain a new action
\begin{equation}
    S_1 = \frac{\b}{2} \sum_p \sum_{i=1}^{N-1} (f^i_p + 2 \pi n^i_p)^2 - \ii \sum_\l \sum_{i=1}^{N-1} v_\l a^i_\l + \frac{1}{2 \b} \sum_p k_p^2 - \ii \sum_p k_p f^N_p - \ii \sum_\l v_l a^N_\l,
\end{equation}
where we have introduced integer-valued plaquette-fields $n^i_p$ ($i=1,\ldots,N-1$) and $k_p$, over which we must perform the summation in the partition function. 
Then exact integration over $a_\ell^N$ gives the constraint
\begin{equation}
    v=-\diff^\dagger k, 
\end{equation}
and then the action becomes 
\begin{eqnarray}
S_2 &=& \frac{\b}{2} \sum_p \sum_{i=1}^{N-1} (f^i_p + 2 \pi n^i_p)^2 + \ii \sum_\l \sum_{i=1}^{N-1} (\dd^\dagger k)_\ell a^i_\ell + \frac{1}{2 \b} \sum_p k^2_p\nonumber\\
&=&\frac{\b}{2} \sum_p \sum_{i=1}^{N-1} (f^i_p + 2 \pi n^i_p)^2 + \ii \sum_\l \sum_{i=1}^{N-1} k_p f^i_p + \frac{1}{2 \b} \sum_p k^2_p.
\end{eqnarray}
Applying the Poisson summation formula in terms of $k_p$, we get 
\begin{equation}
    S_3 = \frac{\b}{2} \sum_p \sum_{i=1}^{N-1} (f^i_p + 2 \pi n^i_p)^2 + {\b\over 2} \sum_p \left(\sum_{i=1}^{N-1} f^i_p + 2 \pi n_p \right)^2,
\end{equation}
where a new integer-valued plaquette-field $n_p$ has taken the place of $k_p$.

For convenience, let us rewrite this last action in the form
\begin{equation}
    S_3 = \frac{\b}{2} \sum_{i=1}^{N-1} (f^i_p + b^i_p)^2 + \frac{\b}{2} \left( \sum_{i=1}^{N-1} f^i_p + b_p \right)^2
\end{equation}
by defining $b^i_p \equiv 2 \pi n^i_p$, $b_p \equiv 2 \pi n_p$. Completing the square then yields
\begin{equation}
    S_3 =  \frac{\b}{2} \sum_p \sum_{i,j=1}^{N-1} D^{ij}(f^i_p + b^i_p + w_p/N) (f^j_p + b^j_p + w_p/N) + \frac{\b}{2} \sum_p w_p^2,
\end{equation}
where we have defined 
\begin{equation}
    w_p \equiv b_p - \sum_{i=1}^{N-1} b^i_p \,, \qquad D^{ij} \equiv 1 + \delta^{i,j}.
\end{equation}
At this point, we realize that if we are only interested in the weak coupling regime, then since the fluctuations in $w_p$ are gapped and discrete, we are entitled to set $w_p=0$. Making this step leaves us with the action
\begin{equation}
    S_4 = \frac{\b}{2} \sum_p \sum_{i,j=1}^{N-1} D^{ij}(f^i_p + b^i_p) (f^j_p + b^j_p).
\end{equation}
Now we note that the unimodular matrix
\begin{equation}
    M^{ij} = \delta^{i,j} - \delta^{i+1,j}
\end{equation}
satisfies
\begin{equation}
    M^{\rm t} D M = C,
\end{equation}
where $C$ is the Cartan matrix of ${\rm SU}(N)$:
\begin{equation}
    C^{ij} = \bm \alpha^i \cdot \bm \alpha^j = 2 \delta^{i,j} - \delta^{i,j+1} - \delta^{i+1,j}.
\end{equation}
It follows that we can make the field redefinitions
\begin{equation}
    a^i_\l \rightarrow \sum_{j=1}^{N-1} M^{ij} A^j_\l \,, \qquad b^i_p \rightarrow \sum_{j=1}^{N-1} M^{ij} B^j_p
\end{equation}
with $A^i_\l \in [0,2\pi]$, $B^i_p \in 2 \pi \Z$, by the unimodularity of $M$. This yields
\begin{equation}
    S = \frac{\b}{2} \sum_p \sum_{i,j=1}^{N-1} C^{ij} (F^i_p+B^i_p) (F^j_p+B^j_p)
\end{equation}
which is equivalent to \eqref{multiVillain}.

\bibliographystyle{JHEP}
\bibliography{QFT, QFT-Mithat} 

\providecommand{\href}[2]{#2}\begingroup\raggedright\begin{thebibliography}{10}

\bibitem{Polyakov:1975rs}
A.~M. Polyakov, \emph{{Compact Gauge Fields and the Infrared Catastrophe}},
  \href{http://dx.doi.org/10.1016/0370-2693(75)90162-8}{\emph{Phys. Lett.} {\bf
  B59} (1975) 82--84}.

\bibitem{Polyakov:1987ez}
A.~Polyakov, \emph{Gauge Fields and Strings (Mathematical Reports,)}.
\newblock CRC Press, 1~ed., 9, 1987.

\bibitem{Seiberg:1994rs}
N.~Seiberg and E.~Witten, \emph{{Electric - magnetic duality, monopole
  condensation, and confinement in N=2 supersymmetric Yang-Mills theory}},
  \href{http://dx.doi.org/10.1016/0550-3213(94)90124-4}{\emph{Nucl. Phys.} {\bf
  B426} (1994) 19--52}, [\href{https://arxiv.org/abs/hep-th/9407087}{{\tt
  hep-th/9407087}}].

\bibitem{Unsal:2008ch}
M.~Unsal and L.~G. Yaffe, \emph{{Center-stabilized Yang-Mills theory:
  Confinement and large N volume independence}},
  \href{http://dx.doi.org/10.1103/PhysRevD.78.065035}{\emph{Phys. Rev.} {\bf
  D78} (2008) 065035}, [\href{https://arxiv.org/abs/0803.0344}{{\tt
  0803.0344}}].

\bibitem{Unsal:2007jx}
M.~Unsal, \emph{{Magnetic bion condensation: A New mechanism of confinement and
  mass gap in four dimensions}},
  \href{http://dx.doi.org/10.1103/PhysRevD.80.065001}{\emph{Phys. Rev.} {\bf
  D80} (2009) 065001}, [\href{https://arxiv.org/abs/0709.3269}{{\tt
  0709.3269}}].

\bibitem{Douglas:1995nw}
M.~R. Douglas and S.~H. Shenker, \emph{{Dynamics of SU(N) supersymmetric gauge
  theory}}, \href{http://dx.doi.org/10.1016/0550-3213(95)00258-T}{\emph{Nucl.
  Phys. B} {\bf 447} (1995) 271--296},
  [\href{https://arxiv.org/abs/hep-th/9503163}{{\tt hep-th/9503163}}].

\bibitem{Argyres:1994xh}
P.~C. Argyres and A.~E. Faraggi, \emph{{The vacuum structure and spectrum of
  N=2 supersymmetric SU(n) gauge theory}},
  \href{http://dx.doi.org/10.1103/PhysRevLett.74.3931}{\emph{Phys. Rev. Lett.}
  {\bf 74} (1995) 3931--3934},
  [\href{https://arxiv.org/abs/hep-th/9411057}{{\tt hep-th/9411057}}].

\bibitem{Klemm:1994qj}
A.~Klemm, W.~Lerche, S.~Yankielowicz and S.~Theisen, \emph{{On the monodromies
  of N=2 supersymmetric Yang-Mills theory}},
  \href{https://arxiv.org/abs/hep-th/9412158}{{\tt hep-th/9412158}}.

\bibitem{Poppitz:2017ivi}
E.~Poppitz and M.~Shalchian~T., \emph{{String tensions in deformed Yang-Mills
  theory}}, \href{http://dx.doi.org/10.1007/JHEP01(2018)029}{\emph{JHEP} {\bf
  01} (2018) 029}, [\href{https://arxiv.org/abs/1708.08821}{{\tt 1708.08821}}].

\bibitem{Ambjorn:1984dp}
J.~Ambjorn, P.~Olesen and C.~Peterson, \emph{{Stochastic Confinement and
  Dimensional Reduction. 2. Three-dimensional SU(2) Lattice Gauge Theory}},
  \href{http://dx.doi.org/10.1016/0550-3213(84)90242-6}{\emph{Nucl. Phys. B}
  {\bf 240} (1984) 533--542}.

\bibitem{Poulis:1995nn}
G.~I. Poulis and H.~D. Trottier, \emph{{'Gluelump' spectrum and adjoint source
  potential in lattice QCD in three-dimensions}},
  \href{http://dx.doi.org/10.1016/S0370-2693(97)88182-8}{\emph{Phys. Lett. B}
  {\bf 400} (1997) 358--363},
  [\href{https://arxiv.org/abs/hep-lat/9504015}{{\tt hep-lat/9504015}}].

\bibitem{Bali:2000un}
G.~S. Bali, \emph{{Casimir scaling of SU(3) static potentials}},
  \href{http://dx.doi.org/10.1103/PhysRevD.62.114503}{\emph{Phys. Rev. D} {\bf
  62} (2000) 114503}, [\href{https://arxiv.org/abs/hep-lat/0006022}{{\tt
  hep-lat/0006022}}].

\bibitem{Philipsen:1999wf}
O.~Philipsen and H.~Wittig, \emph{{String breaking in SU(2) Yang-Mills theory
  with adjoint sources}},
  \href{http://dx.doi.org/10.1016/S0370-2693(99)00183-5}{\emph{Phys. Lett. B}
  {\bf 451} (1999) 146--154},
  [\href{https://arxiv.org/abs/hep-lat/9902003}{{\tt hep-lat/9902003}}].

\bibitem{Stephenson:1999kh}
P.~Stephenson, \emph{{Breaking of the adjoint string in (2+1)-dimensions}},
  \href{http://dx.doi.org/10.1016/S0550-3213(99)00210-2}{\emph{Nucl. Phys. B}
  {\bf 550} (1999) 427--448},
  [\href{https://arxiv.org/abs/hep-lat/9902002}{{\tt hep-lat/9902002}}].

\bibitem{deForcrand:1999kr}
P.~de~Forcrand and O.~Philipsen, \emph{{Adjoint string breaking in 4-D SU(2)
  Yang-Mills theory}},
  \href{http://dx.doi.org/10.1016/S0370-2693(00)00117-9}{\emph{Phys. Lett. B}
  {\bf 475} (2000) 280--288},
  [\href{https://arxiv.org/abs/hep-lat/9912050}{{\tt hep-lat/9912050}}].

\bibitem{Greensite:2003bk}
J.~Greensite, \emph{{The Confinement problem in lattice gauge theory}},
  \href{http://dx.doi.org/10.1016/S0146-6410(03)90012-3}{\emph{Prog. Part.
  Nucl. Phys.} {\bf 51} (2003) 1},
  [\href{https://arxiv.org/abs/hep-lat/0301023}{{\tt hep-lat/0301023}}].

\bibitem{Wellegehausen:2010ai}
B.~H. Wellegehausen, A.~Wipf and C.~Wozar, \emph{{Casimir Scaling and String
  Breaking in G(2) Gluodynamics}},
  \href{http://dx.doi.org/10.1103/PhysRevD.83.016001}{\emph{Phys. Rev. D} {\bf
  83} (2011) 016001}, [\href{https://arxiv.org/abs/1006.2305}{{\tt
  1006.2305}}].

\bibitem{Gaiotto:2014kfa}
D.~Gaiotto, A.~Kapustin, N.~Seiberg and B.~Willett, \emph{{Generalized Global
  Symmetries}}, \href{http://dx.doi.org/10.1007/JHEP02(2015)172}{\emph{JHEP}
  {\bf 02} (2015) 172}, [\href{https://arxiv.org/abs/1412.5148}{{\tt
  1412.5148}}].

\bibitem{Sharpe:2015mja}
E.~Sharpe, \emph{{Notes on generalized global symmetries in QFT}},
  \href{http://dx.doi.org/10.1002/prop.201500048}{\emph{Fortsch. Phys.} {\bf
  63} (2015) 659--682}, [\href{https://arxiv.org/abs/1508.04770}{{\tt
  1508.04770}}].

\bibitem{Kapustin:2013uxa}
A.~Kapustin and R.~Thorngren, \emph{{Higher symmetry and gapped phases of gauge
  theories}},  \href{https://arxiv.org/abs/1309.4721}{{\tt 1309.4721}}.

\bibitem{Cordova:2018cvg}
C.~Cordova, T.~T. Dumitrescu and K.~Intriligator, \emph{{Exploring 2-Group
  Global Symmetries}},
  \href{http://dx.doi.org/10.1007/JHEP02(2019)184}{\emph{JHEP} {\bf 02} (2019)
  184}, [\href{https://arxiv.org/abs/1802.04790}{{\tt 1802.04790}}].

\bibitem{Benini:2018reh}
F.~Benini, C.~Cordova and P.-S. Hsin, \emph{{On 2-Group Global Symmetries and
  Their Anomalies}},  \href{https://arxiv.org/abs/1803.09336}{{\tt
  1803.09336}}.

\bibitem{Misumi:2019dwq}
T.~Misumi, Y.~Tanizaki and M.~\"Unsal, \emph{{Fractional $\theta$ angle, 't
  Hooft anomaly, and quantum instantons in charge-$q$ multi-flavor Schwinger
  model}}, \href{http://dx.doi.org/10.1007/JHEP07(2019)018}{\emph{JHEP} {\bf
  07} (2019) 018}, [\href{https://arxiv.org/abs/1905.05781}{{\tt 1905.05781}}].

\bibitem{Tanizaki:2019rbk}
Y.~Tanizaki and M.~Unsal, \emph{{Modified instanton sum in QCD and
  higher-groups}}, \href{http://dx.doi.org/10.1007/JHEP03(2020)123}{\emph{JHEP}
  {\bf 03} (2020) 123}, [\href{https://arxiv.org/abs/1912.01033}{{\tt
  1912.01033}}].

\bibitem{Hidaka:2020iaz}
Y.~Hidaka, M.~Nitta and R.~Yokokura, \emph{{Higher-form symmetries and 3-group
  in axion electrodynamics}},
  \href{http://dx.doi.org/10.1016/j.physletb.2020.135672}{\emph{Phys. Lett. B}
  {\bf 808} (2020) 135672}, [\href{https://arxiv.org/abs/2006.12532}{{\tt
  2006.12532}}].

\bibitem{Hidaka:2020izy}
Y.~Hidaka, M.~Nitta and R.~Yokokura, \emph{{Global 3-group symmetry and 't
  Hooft anomalies in axion electrodynamics}},
  \href{https://arxiv.org/abs/2009.14368}{{\tt 2009.14368}}.

\bibitem{Bhardwaj:2017xup}
L.~Bhardwaj and Y.~Tachikawa, \emph{{On finite symmetries and their gauging in
  two dimensions}},
  \href{http://dx.doi.org/10.1007/JHEP03(2018)189}{\emph{JHEP} {\bf 03} (2018)
  189}, [\href{https://arxiv.org/abs/1704.02330}{{\tt 1704.02330}}].

\bibitem{Buican:2017rxc}
M.~Buican and A.~Gromov, \emph{{Anyonic Chains, Topological Defects, and
  Conformal Field Theory}},
  \href{http://dx.doi.org/10.1007/s00220-017-2995-6}{\emph{Commun. Math. Phys.}
  {\bf 356} (2017) 1017--1056}, [\href{https://arxiv.org/abs/1701.02800}{{\tt
  1701.02800}}].

\bibitem{Freed:2018cec}
D.~S. Freed and C.~Teleman, \emph{{Topological dualities in the Ising model}},
  \href{https://arxiv.org/abs/1806.00008}{{\tt 1806.00008}}.

\bibitem{Chang:2018iay}
C.-M. Chang, Y.-H. Lin, S.-H. Shao, Y.~Wang and X.~Yin, \emph{{Topological
  Defect Lines and Renormalization Group Flows in Two Dimensions}},
  \href{http://dx.doi.org/10.1007/JHEP01(2019)026}{\emph{JHEP} {\bf 01} (2019)
  026}, [\href{https://arxiv.org/abs/1802.04445}{{\tt 1802.04445}}].

\bibitem{Thorngren:2019iar}
R.~Thorngren and Y.~Wang, \emph{{Fusion Category Symmetry I: Anomaly In-Flow
  and Gapped Phases}},  \href{https://arxiv.org/abs/1912.02817}{{\tt
  1912.02817}}.

\bibitem{Ji:2019jhk}
W.~Ji and X.-G. Wen, \emph{{Categorical symmetry and noninvertible anomaly in
  symmetry-breaking and topological phase transitions}},
  \href{http://dx.doi.org/10.1103/PhysRevResearch.2.033417}{\emph{Phys. Rev.
  Res.} {\bf 2} (2020) 033417}, [\href{https://arxiv.org/abs/1912.13492}{{\tt
  1912.13492}}].

\bibitem{Rudelius:2020orz}
T.~Rudelius and S.-H. Shao, \emph{{Topological Operators and Completeness of
  Spectrum in Discrete Gauge Theories}},
  \href{https://arxiv.org/abs/2006.10052}{{\tt 2006.10052}}.

\bibitem{Komargodski:2020mxz}
Z.~Komargodski, K.~Ohmori, K.~Roumpedakis and S.~Seifnashri, \emph{{Symmetries
  and Strings of Adjoint QCD${}_2$}},
  \href{https://arxiv.org/abs/2008.07567}{{\tt 2008.07567}}.

\bibitem{Aasen:2020jwb}
D.~Aasen, P.~Fendley and R.~S. Mong, \emph{{Topological Defects on the Lattice:
  Dualities and Degeneracies}},  \href{https://arxiv.org/abs/2008.08598}{{\tt
  2008.08598}}.

\bibitem{Wilson:1974sk}
K.~G. Wilson, \emph{{Confinement of Quarks}},
  \href{http://dx.doi.org/10.1103/PhysRevD.10.2445}{\emph{Phys. Rev.} {\bf D10}
  (1974) 2445--2459}.

\bibitem{Villain:1974ir}
J.~Villain, \emph{Theory of one- and two-dimensional magnets with an easy
  magnetization plane. ii. the planar, classical, two-dimensional magnet},
  \href{http://dx.doi.org/10.1051/jphys:01975003606058100}{\emph{J. Phys.
  France} {\bf 36} (1975) 581--590}.

\bibitem{Sulejmanpasic:2019ytl}
T.~Sulejmanpasic and C.~Gattringer, \emph{{Abelian gauge theories on the
  lattice: $\theta$-terms and compact gauge theory with(out) monopoles}},
  \href{http://dx.doi.org/10.1016/j.nuclphysb.2019.114616}{\emph{Nucl. Phys.}
  {\bf B943} (2019) 114616}, [\href{https://arxiv.org/abs/1901.02637}{{\tt
  1901.02637}}].

\bibitem{Banks:1977cc}
T.~Banks, R.~Myerson and J.~B. Kogut, \emph{{Phase Transitions in Abelian
  Lattice Gauge Theories}},
  \href{http://dx.doi.org/10.1016/0550-3213(77)90129-8}{\emph{Nucl. Phys. B}
  {\bf 129} (1977) 493--510}.

\bibitem{Savit:1977fw}
R.~Savit, \emph{{Topological Excitations in U(1) Invariant Theories}},
  \href{http://dx.doi.org/10.1103/PhysRevLett.39.55}{\emph{Phys. Rev. Lett.}
  {\bf 39} (1977) 55}.

\bibitem{Gopfert:1981er}
M.~Gopfert and G.~Mack, \emph{{Proof of Confinement of Static Quarks in
  Three-Dimensional U(1) Lattice Gauge Theory for All Values of the Coupling
  Constant}}, \href{http://dx.doi.org/10.1007/BF01961240}{\emph{Commun. Math.
  Phys.} {\bf 82} (1981) 545}.

\bibitem{Ukawa:1979yv}
A.~Ukawa, P.~Windey and A.~H. Guth, \emph{{Dual Variables for Lattice Gauge
  Theories and the Phase Structure of Z(N) Systems}},
  \href{http://dx.doi.org/10.1103/PhysRevD.21.1013}{\emph{Phys. Rev. D} {\bf
  21} (1980) 1013}.

\bibitem{Dorey:1999sj}
N.~Dorey, \emph{{An Elliptic superpotential for softly broken N=4
  supersymmetric Yang-Mills theory}},
  \href{http://dx.doi.org/10.1088/1126-6708/1999/07/021}{\emph{JHEP} {\bf 07}
  (1999) 021}, [\href{https://arxiv.org/abs/hep-th/9906011}{{\tt
  hep-th/9906011}}].

\bibitem{Bogomolny:1975de}
E.~B. Bogomolny, \emph{{Stability of Classical Solutions}}, {\emph{Sov. J.
  Nucl. Phys.} {\bf 24} (1976) 449}.

\bibitem{Prasad:1975kr}
M.~K. Prasad and C.~M. Sommerfield, \emph{{An Exact Classical Solution for the
  't Hooft Monopole and the Julia-Zee Dyon}},
  \href{http://dx.doi.org/10.1103/PhysRevLett.35.760}{\emph{Phys. Rev. Lett.}
  {\bf 35} (1975) 760--762}.

\bibitem{Anber:2015kea}
M.~M. Anber, E.~Poppitz and T.~Sulejmanpasic, \emph{{Strings from domain walls
  in supersymmetric Yang-Mills theory and adjoint QCD}},
  \href{http://dx.doi.org/10.1103/PhysRevD.92.021701}{\emph{Phys. Rev.} {\bf
  D92} (2015) 021701}, [\href{https://arxiv.org/abs/1501.06773}{{\tt
  1501.06773}}].

\bibitem{Cho:2014jfa}
G.~Y. Cho, J.~C.~Y. Teo and S.~Ryu, \emph{{Conflicting Symmetries in
  Topologically Ordered Surface States of Three-dimensional Bosonic Symmetry
  Protected Topological Phases}},
  \href{http://dx.doi.org/10.1103/PhysRevB.89.235103}{\emph{Phys. Rev.} {\bf
  B89} (2014) 235103}, [\href{https://arxiv.org/abs/1403.2018}{{\tt
  1403.2018}}].

\bibitem{Kapustin:2014lwa}
A.~Kapustin and R.~Thorngren, \emph{{Anomalies of discrete symmetries in three
  dimensions and group cohomology}},
  \href{http://dx.doi.org/10.1103/PhysRevLett.112.231602}{\emph{Phys. Rev.
  Lett.} {\bf 112} (2014) 231602}, [\href{https://arxiv.org/abs/1403.0617}{{\tt
  1403.0617}}].

\bibitem{Kapustin:2014zva}
A.~Kapustin and R.~Thorngren, \emph{{Anomalies of discrete symmetries in
  various dimensions and group cohomology}},
  \href{https://arxiv.org/abs/1404.3230}{{\tt 1404.3230}}.

\bibitem{Dijkgraaf:1990nc}
R.~Dijkgraaf and E.~Witten, \emph{{Mean Field Theory, Topological Field Theory,
  and Multimatrix Models}},
  \href{http://dx.doi.org/10.1016/0550-3213(90)90324-7}{\emph{Nucl.Phys.} {\bf
  B342} (1990) 486--522}.

\bibitem{Gaiotto:2017yup}
D.~Gaiotto, A.~Kapustin, Z.~Komargodski and N.~Seiberg, \emph{{Theta, Time
  Reversal, and Temperature}},
  \href{http://dx.doi.org/10.1007/JHEP05(2017)091}{\emph{JHEP} {\bf 05} (2017)
  091}, [\href{https://arxiv.org/abs/1703.00501}{{\tt 1703.00501}}].

\bibitem{Tanizaki:2017bam}
Y.~Tanizaki and Y.~Kikuchi, \emph{{Vacuum structure of bifundamental gauge
  theories at finite topological angles}},
  \href{http://dx.doi.org/10.1007/JHEP06(2017)102}{\emph{JHEP} {\bf 06} (2017)
  102}, [\href{https://arxiv.org/abs/1705.01949}{{\tt 1705.01949}}].

\bibitem{Komargodski:2017smk}
Z.~Komargodski, T.~Sulejmanpasic and M.~Unsal, \emph{{Walls, anomalies, and
  deconfinement in quantum antiferromagnets}},
  \href{http://dx.doi.org/10.1103/PhysRevB.97.054418}{\emph{Phys. Rev.} {\bf
  B97} (2018) 054418}, [\href{https://arxiv.org/abs/1706.05731}{{\tt
  1706.05731}}].

\bibitem{Shimizu:2017asf}
H.~Shimizu and K.~Yonekura, \emph{{Anomaly constraints on deconfinement and
  chiral phase transition}},
  \href{http://dx.doi.org/10.1103/PhysRevD.97.105011}{\emph{Phys. Rev.} {\bf
  D97} (2018) 105011}, [\href{https://arxiv.org/abs/1706.06104}{{\tt
  1706.06104}}].

\bibitem{Gaiotto:2017tne}
D.~Gaiotto, Z.~Komargodski and N.~Seiberg, \emph{{Time-reversal breaking in
  QCD$_{4}$, walls, and dualities in 2 + 1 dimensions}},
  \href{http://dx.doi.org/10.1007/JHEP01(2018)110}{\emph{JHEP} {\bf 01} (2018)
  110}, [\href{https://arxiv.org/abs/1708.06806}{{\tt 1708.06806}}].

\bibitem{Tanizaki:2017qhf}
Y.~Tanizaki, T.~Misumi and N.~Sakai, \emph{{Circle compactification and 't
  Hooft anomaly}}, \href{http://dx.doi.org/10.1007/JHEP12(2017)056}{\emph{JHEP}
  {\bf 12} (2017) 056}, [\href{https://arxiv.org/abs/1710.08923}{{\tt
  1710.08923}}].

\bibitem{Tanizaki:2017mtm}
Y.~Tanizaki, Y.~Kikuchi, T.~Misumi and N.~Sakai, \emph{{Anomaly matching for
  phase diagram of massless $\mathbb{Z}_N$-QCD}},
  \href{http://dx.doi.org/10.1103/PhysRevD.97.054012}{\emph{Phys. Rev.} {\bf
  D97} (2018) 054012}, [\href{https://arxiv.org/abs/1711.10487}{{\tt
  1711.10487}}].

\bibitem{Tanizaki:2018wtg}
Y.~Tanizaki, \emph{{Anomaly constraint on massless QCD and the role of
  Skyrmions in chiral symmetry breaking}},
  \href{http://dx.doi.org/10.1007/JHEP08(2018)171}{\emph{JHEP} {\bf 08} (2018)
  171}, [\href{https://arxiv.org/abs/1807.07666}{{\tt 1807.07666}}].

\bibitem{Komargodski:2017dmc}
Z.~Komargodski, A.~Sharon, R.~Thorngren and X.~Zhou, \emph{{Comments on Abelian
  Higgs Models and Persistent Order}},
  \href{http://dx.doi.org/10.21468/SciPostPhys.6.1.003}{\emph{SciPost Phys.}
  {\bf 6} (2019) 003}, [\href{https://arxiv.org/abs/1705.04786}{{\tt
  1705.04786}}].

\bibitem{Kikuchi:2017pcp}
Y.~Kikuchi and Y.~Tanizaki, \emph{{Global inconsistency, 't~Hooft anomaly, and
  level crossing in quantum mechanics}},
  \href{http://dx.doi.org/10.1093/ptep/ptx148}{\emph{Prog. Theor. Exp. Phys.}
  {\bf 2017} (2017) 113B05}, [\href{https://arxiv.org/abs/1708.01962}{{\tt
  1708.01962}}].

\bibitem{Tanizaki:2018xto}
Y.~Tanizaki and T.~Sulejmanpasic, \emph{{Anomaly and global inconsistency
  matching: $\theta$-angles, $SU(3)/U(1)^2$ nonlinear sigma model, $SU(3)$
  chains and its generalizations}},
  \href{http://dx.doi.org/10.1103/PhysRevB.98.115126}{\emph{Phys. Rev.} {\bf
  B98} (2018) 115126}, [\href{https://arxiv.org/abs/1805.11423}{{\tt
  1805.11423}}].

\bibitem{Karasik:2019bxn}
A.~Karasik and Z.~Komargodski, \emph{The bi-fundamental gauge theory in 3+1
  dimensions: The vacuum structure and a cascade},
  \href{http://dx.doi.org/10.1007/JHEP05(2019)144}{\emph{JHEP} {\bf 05} (2019)
  144}, [\href{https://arxiv.org/abs/1904.09551}{{\tt 1904.09551}}].

\bibitem{Cordova:2019jnf}
C.~Cordova, D.~S. Freed, H.~T. Lam and N.~Seiberg, \emph{{Anomalies in the
  Space of Coupling Constants and Their Dynamical Applications I}},
  \href{http://dx.doi.org/10.21468/SciPostPhys.8.1.001}{\emph{SciPost Phys.}
  {\bf 8} (2020) 001}, [\href{https://arxiv.org/abs/1905.09315}{{\tt
  1905.09315}}].

\bibitem{Creutz:1979kf}
M.~Creutz, L.~Jacobs and C.~Rebbi, \emph{{Experiments with a Gauge Invariant
  Ising System}},
  \href{http://dx.doi.org/10.1103/PhysRevLett.42.1390}{\emph{Phys. Rev. Lett.}
  {\bf 42} (1979) 1390}.

\bibitem{Creutz:1979zg}
M.~Creutz, L.~Jacobs and C.~Rebbi, \emph{{Monte Carlo Study of Abelian Lattice
  Gauge Theories}},
  \href{http://dx.doi.org/10.1103/PhysRevD.20.1915}{\emph{Phys. Rev.} {\bf D20}
  (1979) 1915}.

\bibitem{Cardy:1981qy}
J.~L. Cardy and E.~Rabinovici, \emph{{Phase Structure of Z(p) Models in the
  Presence of a Theta Parameter}},
  \href{http://dx.doi.org/10.1016/0550-3213(82)90463-1}{\emph{Nucl. Phys.} {\bf
  B205} (1982) 1--16}.

\bibitem{Cardy:1981fd}
J.~L. Cardy, \emph{{Duality and the Theta Parameter in Abelian Lattice
  Models}}, \href{http://dx.doi.org/10.1016/0550-3213(82)90464-3}{\emph{Nucl.
  Phys.} {\bf B205} (1982) 17--26}.

\bibitem{Honda:2020txe}
M.~Honda and Y.~Tanizaki, \emph{{Topological aspects of $4$D Abelian lattice
  gauge theories with the $\theta$ parameter}},
  \href{http://dx.doi.org/10.1007/JHEP12(2020)154}{\emph{JHEP} {\bf 12} (2020)
  154}, [\href{https://arxiv.org/abs/2009.10183}{{\tt 2009.10183}}].

\end{thebibliography}\endgroup

\end{document}